\documentclass[transaction]{IEEEtran}
\usepackage{graphicx}
\usepackage{epstopdf}
\usepackage[latin1]{inputenc}
\usepackage{amsmath,amssymb,amscd,latexsym,dsfont}
\usepackage[english]{babel}
\usepackage{tabularx,cite}
\usepackage{graphicx}
\usepackage{algpseudocode}
\usepackage{algorithm}
\usepackage{algorithmicx}
\usepackage{float}
\usepackage{multicol}
\usepackage{psfrag}
\usepackage{comment,psfrag,subfigure,enumerate}
\usepackage{color}
\usepackage{amsthm}

\ifCLASSINFOpdf
\else
\fi
\begin{document}
%
\title{Green Communication via Power-optimized HARQ Protocols}
\author{\IEEEauthorblockN{Behrooz Makki, Alexandre Graell i Amat, \emph{Senior Member, IEEE} and Thomas Eriksson}\\
\thanks{Behrooz Makki, Alexandre Graell i Amat and Thomas Eriksson are with Department of Signals and Systems,
Chalmers University of Technology, Gothenburg, Sweden, Email: \{behrooz.makki, alexandre.graell, thomase\}@chalmers.se}
\thanks{Alexandre Graell i Amat was supported by the Swedish Agency for Innovation Systems (VINNOVA) under the P36604-1 MAGIC project.}
}
%
\maketitle
\vspace{-0mm}
\begin{abstract}
Recently, efficient use of energy has become an essential research topic for green communication. This paper studies the effect of optimal power controllers on the performance of delay-sensitive communication setups utilizing hybrid automatic repeat request (HARQ). The results are obtained for repetition time diversity (RTD) and incremental redundancy (INR) HARQ protocols. In all cases, the optimal power allocation, minimizing the outage-limited average transmission power, is obtained under both continuous and bursting communication models. Also, we investigate the system throughput in different conditions. The results indicate that the power efficiency is increased substantially, if adaptive power allocation is utilized. For instance, assume Rayleigh-fading channel, a maximum of two (re)transmission rounds with rates $\{1,\frac{1}{2}\}$ nats-per-channel-use and an outage probability constraint ${10}^{-3}$. Then, compared to uniform power allocation, optimal power allocation in RTD reduces the average power by 9 and 11 dB in the bursting and continuous communication models, respectively. In INR, these values are obtained to be 8 and 9 dB, respectively.
\end{abstract}
%
\IEEEpeerreviewmaketitle
\vspace{-0mm}
\section{Introduction}
\vspace{-0mm}

The first and the most important requirement in many wireless applications is that the data must be decodable at the receiver. This problem is often studied under the topic of \emph{outage-limited} data transmission  \cite{isitakhodemun,excellentref,5711682,ekbatanioutage,outageHARQ}; An outage happens when the data can not be decoded at the receiver. With an outage constraint, the data transmission is successful if the codewords are decodable, regardless of whether they lead to maximum system throughput or not.


Assuming delay-sensitive communication over block-fading channels, it is well-accepted that infinite power is required to achieve zero outage probability for all channel conditions \cite{isitakhodemun,5711682,ekbatanioutage,excellentref,outageHARQ}. This is because the channel may fall in deep fading conditions and so infinitely large powers are needed for \emph{channel inversion}, i.e., for compensating the channel \emph{bad} conditions. In these cases, the strict outage requirements are replaced by more relaxed probabilistic constraints \cite{isitakhodemun,ekbatanioutage,outageHARQ,excellentref,5711682}, where ``a service is acceptable as long as the data is always decoded with some probability $\epsilon$." Here, $\epsilon$ is a parameter representing the system outage tolerance.

Due to the fast growth of wireless networks and the data-intensive applications produced by smart phones, green communication via improving the power efficiency is becoming increasingly important for wireless communication. The network data volume is expected to increase by a factor of $2$ every year, associated with $16-20\%$ increase of energy consumption, which contributes about $2\%$ of global $CO_2$ emissions \cite{Gartner}. Hence, minimizing the power consumption is a very important design consideration, and power-efficient data transmission schemes must be taken into account for the wireless networks \cite{1321221}. Green radio has thereby been proposed as an effective solution and is becoming the mainstream for future wireless network design \cite{5783982,greenref3,greenref4,greenref5,greenref6,greenref1,greenref2}.

From another perspective, hybrid automatic repeat request (HARQ) is a well-known approach applied in today's wireless networks to increase the data transmission reliability and efficiency \cite{5961851,pphdexcelent,outageHARQ,letterkhodemun,5336856,cairearq1,cairearq2,ARQ20112,ARQ20113,sag5771499,ARQGlarsson,1200407,Arulselvan,4356994,5497972,4200959}. The main idea behind the HARQ techniques is to reduce the data outage probability or increase the throughput by retransmitting the data that underwent \emph{bad} channel conditions. Consequently, it is expected that joint implementation of adaptive power controllers and HARQ protocols improve the power efficiency of outage-limited communication systems.

Outage-limited power allocation is an interesting problem which was previously studied by, e.g., Caire \emph{et. al.}, \cite{excellentref} under perfect channel state information (CSI) assumption. Also, \cite{5711682,ekbatanioutage} investigated the same problem in the presence of quantized CSI. In \cite{outageHARQ}, Wu and Jindal studied the outage-limited performance of HARQ protocols in block-fading channels. Here, the results were obtained with no power adaptation and under the assumption that the channel changes in each retransmission round. The outage-limited power allocation problem for the repetition time diversity (RTD) and fixed-length coding incremental redundancy (INR) HARQ protocols were investigated in \cite{sag5771499} and \cite{ARQGlarsson}, respectively. Finally, assuming that the channel changes in each retransmission, \cite{4200959} found the optimal retransmission power for basic automatic repeat request (ARQ) protocols where it was shown that the power should be increasing in every retransmission.

Power allocation in HARQ schemes has also been considered in \cite{5961851} and \cite{5336856}, where the power-limited throughput optimization problem of different HARQ protocols was studied. In \cite{1200407}, the optimal HARQ-based power allocation for minimizing the required number of retransmission rounds was obtained in a down-link wideband code division multiple access (WCDMA) system. Assuming partial CSI at the transmitter, \cite{Arulselvan} and \cite{4356994} studied the power allocation problem in HARQ systems when the power is changed according to the received CSI and by a linear programming approach under a buffer cost constraint, respectively. Finally, \cite{5497972} investigated the power allocation between the source and the relay in a relay channel where, while the powers are fixed in the retransmission rounds, there is a sum power constraint on the relay and the transmitter.

This paper investigates the power allocation problem in HARQ-based systems under different outage probability constraints.
Utilizing different HARQ protocols, the goal is to determine the optimal power controllers satisfying different outage probability constraints. Also, the same setup is valid for the inverse problem where for a given transmission power, the goal is to minimize the outage probability. The results are obtained for the RTD and the INR protocols, under both continuous and bursting communication models and with the average power definition given in, e.g., \cite{excellentref,5281736,ekbatanioutage,33,letterkhodemun,5336856,5961851,Tcomkhodemun}.

Along with the outage probability, the long-term throughput is another metric demonstrating the system long-term performance. We obtain the long-term throughput and give comparisons between the outage-limited average power and throughput of the RTD and INR protocols.

The outage-limited power allocation problem setup of the paper has not been investigated in \cite{5336856,5961851,outageHARQ,1200407,Arulselvan,4356994,5497972,ARQ20112,ARQ20113}. Also, the paper is different from \cite{4200959} as 1) in contrast to \cite{4200959} that deals with basic ARQ, we study the HARQ protocols and 2) the results are obtained under different block-fading channel assumptions. As mentioned before, \cite{sag5771499,ARQGlarsson} have also studied the outage-limited power allocation problem in HARQ protocols. However, there are fundamental differences between the definition of the average power in \cite{sag5771499,ARQGlarsson} and the definition considered in our work. In particular, the definition of power used in \cite{sag5771499,ARQGlarsson} corresponds to the definition of the energy (although in these papers it is referred as power), while in this paper we consider the more common definition of the average power, as in, e.g., \cite{excellentref,5281736,ekbatanioutage,33,letterkhodemun,5336856,5961851,Tcomkhodemun}. The difference in the definition of the average power makes the problem solved in this paper completely different from the one addressed in \cite{sag5771499,ARQGlarsson}, leading to different analytical and numerical results, as well as to different conclusions. Also, some other contributions of our work compared to \cite{sag5771499,ARQGlarsson} (and also \cite{5336856,5961851,outageHARQ,1200407,Arulselvan,4356994,5497972,4200959,ARQ20112,ARQ20113}) are: 1) solving the problem for both continuous and bursting communication models, 2) focusing on variable-length INR HARQ, instead of a fixed-length coding scheme, and 3) studying the long-term throughput in each case.

The main idea of the paper can be explained as follows. With an outage probability constraint, the initial transmission(s) of the HARQ scheme is set to have a small power. If the channel is \emph{bad} the data can not be decoded and is retransmitted with higher powers. On the other hand, if the channel experiences good conditions, this gambling brings high return. With this scheme, which is shown to be optimal in terms of outage-limited average transmission power, the HARQ helps us to exploit the good channel conditions properly, and decrease the average transmission power for a given outage probability constraint. Moreover, we illustrate power adaptation schemes based on reinforcement algorithms which, depending on the fading condition, can lead to performance improvement in the HARQ protocols.

Minimizing the average transmission power with an outage probability constraint, our results show that:
1) for the RTD HARQ protocol, the transmission powers must be increasing in every retransmission. 2) Also, for the INR HARQ protocol, the transmission energies must increase in every retransmission. 3) For sufficiently large number of retransmissions and independent of the fading distribution, the optimal retransmission powers of the RTD and fixed-length coding INR schemes can be determined via a specific sequence of numbers converging to a geometric sequence. Note that these results are in contrast with the results in \cite{sag5771499,ARQGlarsson}, which is due to differences in the objective function of our work and \cite{sag5771499,ARQGlarsson} \footnote{In \cite{sag5771499,ARQGlarsson}, where the goal is to minimize the transmission energy, the optimal (re)transmission powers are found to be neither increasing nor decreasing.}.
4) Finally, optimal power allocation leads to considerable performance improvement in HARQ protocols. For instance, assume Rayleigh-fading channels, a maximum of two (re)transmission rounds with rates $\{1,\frac{1}{2}\}$ nats-per-channel-use (npcu)\footnote{A nat is a unit of information, based on the natural logarithm \cite{4444444444,letterkhodemun,5336856,Tcomkhodemun}. The results can be mapped to the bit unit, if the logarithmic terms are presented in base 2.} and an outage probability constraint ${10}^{-3}$. Then, compared to uniform power allocation, optimal power allocation in the RTD HARQ scheme reduces the average power by 9 and 11 dB in the bursting and continuous communication models, respectively. In the INR, these values are obtained to be 8 and 9dB, respectively. Also, depending on the channel condition, the power efficiency can be improved by reinforcement-based power allocation schemes, at the cost of higher implementation complexity and sensitivity to imperfect feedback signal.

%
\vspace{-0mm}
\section{System model}
\begin{figure}
\vspace{-0mm}
\centering
  \includegraphics[width=1\columnwidth]{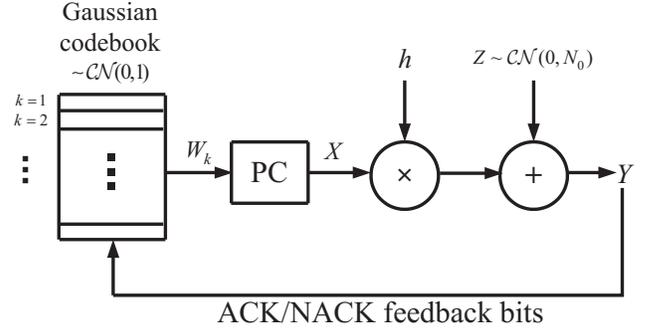}\\
\caption{System model. The codewords are constructed according to the standard Gaussian codes. Therefore, we have $\mathop {\lim }\limits_{L \to \infty } \frac{1}{L}\sum_{i = 1}^L {|{W_k}[i]{|^2}}  \le 1,\, \forall k,$ where $L$ is the length of the codewords. Both fixed-length (RTD) HARQ and variable-length (INR) HARQ schemes are studied in Sections IV and V, respectively. The block PC represents the power controller where the codewords are rescaled based on the number of (re)transmission round. The goal is to determine the optimal power controllers under different outage probability constraints. Finally, the HARQ feedback bits are assumed to be received by the transmitter error- and delay-free. }\label{figure111}
\vspace{-4mm}
\end{figure}
\emph{Channel assumptions:} As illustrated in Fig. 1, we consider a block-fading channel where the power-limited input message $X$ multiplied by the random variable $h$ is summed with an i.i.d. complex Gaussian noise $Z \sim \mathcal{CN}(0,N_0)$ resulting in the output
\vspace{-0mm}
\begin{equation}
\vspace{-0mm}
Y = h X + Z.
\vspace{-0mm}
\end{equation}

Let us define $g=|h|^2$ as the \emph{channel gain} random variable. The channel gain remains constant for a duration of $L_\text{c}$ channel uses (cu)\footnote{To be in harmony with the literature, e.g., \cite{letterkhodemun,5336856,Tcomkhodemun,6006606}, channel use (cu) is considered as the unit for all temporal parameters, including the channel coherence time, the length of a fading block and the codewords length. For every given duration of a channel use, the results can be mapped to other time units, e.g., seconds, via a scaling factor.}, generally determined by the channel coherence time, and then changes independently according to the fading probability density function (pdf) ${f_G}(g)$. Furthermore, with no loss of generality, we consider $N_0 = 1$.

It is assumed that there is perfect instantaneous knowledge about the channel gain at the receiver, which is an acceptable assumption under block-fading conditions \cite{6082492,Karmokar,4200959,isitakhodemun,ekbatanioutage,5711682,excellentref,33,5281736}. Also,
the fading is assumed to be constant over the transmission of one packet, where a packet is defined as the transmission of a codeword along with all its possible retransmissions. This model has been considered in many other papers, e.g. \cite{5281736,33,letterkhodemun,ARQGlarsson,sag5771499,excellentref,ekbatanioutage,5336856,Tcomkhodemun}, and is a good model for stationary or slow-moving users\footnote{As discussed in, e.g., \cite{excellentref}, the information theoretical results of block-fading channels match the results of actual codes for practical code lengths, e.g., $\simeq 100$ channel uses, and provide appropriate performance bounds for systems with smaller code lengths. Also, please see \cite{cooperationlimit,5567190,6093903,720551} for mappings between the block- and continuous-fading channels (With continuous-fading the channel changes in each channel use).}. Also, it corresponds to the worst case, since no time diversity can be exploited if the channel is fixed within a packet period. Later, in Sections VII and VIII, we relax the block-fading assumption and extend the results to the case with fast-fading conditions. All results are presented in natural logarithm basis and, as each transmission experiences an AWGN channel, the results are restricted to Gaussian input distributions. Finally, the main focus is on two stop-and-wait HARQ protocols:
\begin{itemize}
\item[1)] \textbf{Repetition time diversity}. This scheme belongs to the \emph{diversity combining} category of HARQ protocols \cite{5961851,sag5771499,5336856,6006606} where the same data is repeated in the (re)transmission rounds and, in each round, the receiver performs maximum ratio combining of all received signals.
\item[2)] \textbf{Incremental Redundancy}. The INR belongs to the category of \emph{code combining} protocols \cite{5961851,5336856,6006606,letterkhodemun,ARQGlarsson}. Here, a codeword is sent with an aggressive rate in the first round. Then, if the receiver cannot decode the initial codeword, further parity bits are sent in the next retransmission rounds and in each round the receiver decodes the data based on all received signals.
\end{itemize}

\emph{Objective function:} The main goal of this paper is to design the optimal power controllers, i.e., the block PC in Fig. 1, such that a given outage probability constraint is satisfied with minimum average transmission power. The codebooks are assumed to be constructed by the standard complex Gaussian codes, which have been shown to be optimal for power-limited data transmission in AWGN channels with long codewords \cite{4444444444}. That is, each codeword is constructed according to $\mathcal{CN}(0,1)$. Then, the codewords are rescaled to have different powers, determined based on the number of (re)transmission round. We assume the codewords lengths and rates to be previously designed based on, e.g., data structure and coding complexity. Therefore, the transmission rates are assumed to be out of our control throughout the paper. However, the results are valid for any given codewords rates and number of retransmissions. Hence, this assumption does not affect the generality of the arguments. A number of cases in which the problem setup is applicable are explained in the following.


In general, the problem formulation used here is appropriate for fixed-rate delay-sensitive applications such as voice over internet protocol (VoIP) and fixed-rate video streaming applications \cite{5711682}. The setup becomes more interesting when we remember that many practical ARQ schemes are designed to operate at fixed (re)transmission rates \cite{6006606}. Also, the power control strategy resulting from this problem formulation attempts to fix the transmission rate by compensating for the channel fading and in this sense resembles the power control mechanisms used in WCDMA system standards \cite{5711682,wcdmaref}.
\vspace{-0mm}
\section{Communication models and definitions}
\begin{figure}
\vspace{-0mm}
\centering
  \includegraphics[width=1\columnwidth]{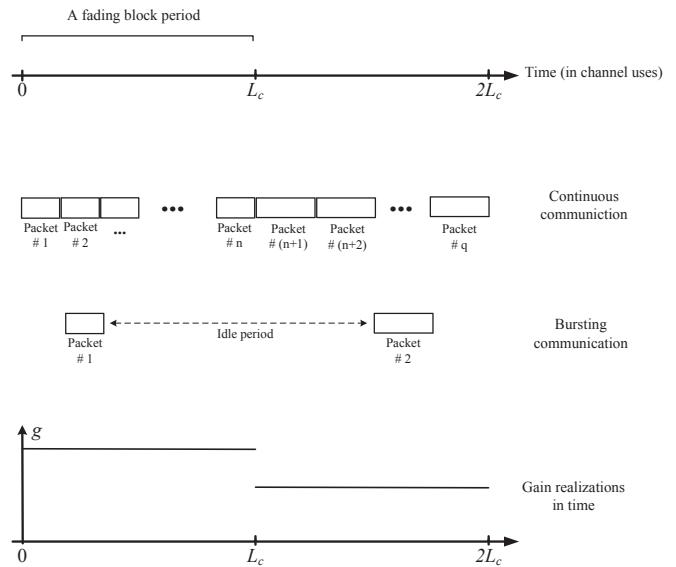}\\
\caption{Description of different data communication models. With bursting (continuous) communication model, one (many) packet(s) is sent within each fading block. A packet is defined as the transmission of a codeword along with all its possible retransmission rounds.}\label{figure111}
\vspace{-3mm}
\end{figure}
We consider both continuous and bursting communication schemes, as illustrated in Fig. 2. Under the continuous communication model, it is assumed that there is an unlimited amount of information available at the transmitter and it is always active \cite{5281736,letterkhodemun,5336856}. In this way, multiple packets, each packet containing multiple HARQ rounds, are transmitted within one fading block of length $L_\text{c}$. When the channel is good, many packets can be sent within the fading block, while only few can be transmitted within the same period for bad channels.

Under the bursting communication model, on the other hand, it is assumed that there is a long idle period between the packet transmissions. Therefore, while the HARQ retransmission rounds of each packet experience the same gain realization, the channel changes independently from one packet to another. To be more clear, all the $L_\text{c}$ channel uses of a fading block are utilized in the continuous communication model. This is because data is continuously transmitted, regardless of whether it is decoded or not. In the bursting communication model, on the other hand, only one packet is sent within each block that, depending on the channel conditions, can be decoded in different (re)transmission rounds. Therefore, the number of channel uses in each block is different. In the following, the long-term throughput and the long-term average transmission power are defined and derived for the continuous communication model. Then, all arguments about performance analysis in the bursting communication model are presented in Section VI. Section VIII presents the simulation results for both cases.
\vspace{-0mm}
\subsection{Long-term throughput}
\vspace{-0mm}
The long-term throughput (in nats-per-channel-use) is defined as
\vspace{-0mm}
\begin{align}
\eta  \buildrel\textstyle.\over= \mathop {\lim }\limits_{k \to \infty } \frac{{{Q^{(k)}}}}{{{\tau ^{(k)}}}} = \frac{{E\{\tilde Q\} }}{{E\{\tilde \tau \} }}
\end{align}
where $Q^{(k)}$ and $\tau^{(k)}$ are the total successfully decoded information nats and the total number of channel uses at the end of $k$ data transmission times. Then, $\tilde Q$ and $\tilde \tau$ are the number of successfully decoded information nats and channel uses in each block, respectively, and $E\{.\}$ is the expectation operator \cite{cairearq1}.

Assuming continuous communication, the long-term throughput can be calculated as follows. Let $R(g)$ be the instantaneous data rate of the HARQ approach for a given gain realization $g$. The total number of information nats that can be decoded in each state is obtained by $Q(g)=L_\text{c} R(g)$. Consequently, the long-term throughput is simplified to
\vspace{-0mm}
\begin{align}
\eta  = \frac{{E\{ {L_\text{c}}R(g)\} }}{{{L_\text{c}}}} = E\{ R(g)\}  = \bar R
\end{align}
where $\bar R$ is the channel average rate \cite{letterkhodemun,5336856,averageratedefinition,5281736,6006606}.
\vspace{-6mm}
\subsection{Long-term average transmission power}

The long-term average transmission power is defined as
\vspace{-0mm}
\begin{align}
\varphi = \mathop {\lim }\limits_{k \to \infty } \frac{{{\xi ^{(k)}}}}{{{\tau ^{(k)}}}} = \frac{{E\{ \tilde \xi \} }}{{E\{ \tilde \tau \} }}
\end{align}
where ${{{\xi ^{(k)}}}}$ and $\tilde \xi$ are the total energy consumed in $k$ data transmission times and within a block period, respectively \cite{excellentref,5281736,ekbatanioutage,33,letterkhodemun,5336856,5961851}. We denote by $P(g)$ the transmission power random variable of an HARQ scheme for a channel gain realization $g$. Then, the average transmission power for the continuous communication model is obtained by
\vspace{-0mm}
\begin{align}
\varphi  = \frac{{E\{ {L_c}P(g)\} }}{{ {L_c} }} = E\{ P(g)\}  = \bar P.
\end{align}

In the bursting communication model, on the other hand, the denominator of, e.g., (2) is not constant. Hence, the long-term throughput and the average transmission power should be directly calculated based on (2) and (4), respectively (please see Section VI for more details).
\vspace{-0mm}
\section{RTD HARQ protocols with an outage probability constraint}
Assuming a continuous communication model, this section first studies the optimal power controllers in RTD HARQ-based schemes constrained to different outage probability constraints. Later, the long-term throughput is investigated.

We consider a maximum of $M$ data retransmission rounds, i.e., each codeword is (re)transmitted a maximum of $M+1$ times, determined by the system delay requirements. Implementing the \emph{Gaussian} codes, the original RTD-based codeword is constructed by encoding ${Q}$ information nats into a codeword of length $L,\, {L} \ll {L_\text{c}},$ and rate $R=\frac{Q}{L}$. The codeword is rescaled in each (re)transmission round to have power $P_m,\,m=1,\ldots,M+1$. Representing the codeword transmitted at the $m$-th (re)transmission round by $\{X_m[i],i=1,\ldots,L\}$ we have
\vspace{-0mm}
\begin{align}
{P_m}= \frac{1}{L}\sum_{i = 1}^{{L}} {|{X_m}[i]{|^2}}.
\end{align}
The (re)transmission continues until an acknowledgement (ACK) feedback bit is received by the transmitter or the maximum permitted retransmission rounds are used. At the end of the $m$-th  (re)transmission round, the receiver performs maximum ratio combining of the $m$ received signals. This process effectively increases the received signal-to-noise ratio (SNR) to ${g\sum_{n = 1}^m {{P_n}} }$ and reduces the data rate to $\frac{R}{m}$. Define ${I_m} \buildrel\textstyle.\over= \frac{1}{m}\log (1 + g\sum_{n = 1}^m {{P_n}} )$ as the instantaneous mutual information and ${\Upsilon _m} \buildrel\textstyle.\over= \{ {I_m} \ge \frac{R}{m}\}$ as the event that the instantaneous mutual information exceeds the equivalent transmission rate at the $m$-th (re)transmission round. The data is successfully decoded at the $m$-th retransmission round if 1) the receiver has not decoded the message in the previous (re)transmissions, i.e., ${I_n} < \frac{R}{n}\,\,\forall n < m$, and 2) using the $m$-th retransmission round it can decode the information, that is, ${I_m} \ge \frac{R}{m}$. Then, as ${\Upsilon _m} \subset {\Upsilon _n},n \le m$, we have
\vspace{-0mm}
\begin{align}
\begin{array}{l}
 \Pr \{ {S_m}\}  = \Pr \{ {{\hat \Upsilon}_1},\ldots,{{\hat \Upsilon}_{m - 1}},{\Upsilon_m}\}  \\
  = \Pr \left\{ {\log (1 + g\sum_{n = 1}^{m - 1} {{P_n}} ) < R \le \log (1 + g\sum_{n = 1}^m {{P_n}} )} \right\} \\= {F_G}(\frac{{{e^R} - 1}}{{\sum_{n = 1}^{m - 1} {{P_n}} }}) - {F_G}(\frac{{{e^R} - 1}}{{\sum_{n = 1}^m {{P_n}} }}), \\
 \Pr \{ {{\hat S}_{M + 1}}\}  = \Pr \left\{ {\log (1 + g\sum_{n = 1}^{M + 1} {{P_n}} ) < R} \right\} \\= {F_G}(\frac{{{e^R} - 1}}{{\sum_{n = 1}^{M + 1} {{P_n}} }}).
 \end{array}
\end{align}
Here, $S_m$ is the event that the data is decoded at the end of the $m$-th round, $\hat U$ denotes the complement of the event $U$ and $F_G$ is the channel gain cumulative distribution function (cdf). Consequently, $\Pr \{ {{\hat S}_{M + 1}}\}$ represents the probability that the data is lost while all retransmission rounds have been used.

On the other hand, the average transmission power at the end of the $m$-th (re)transmission round is
\vspace{-0mm}
\begin{align}
{P^{(m)}} =  \frac{1}{{m{L}}}\sum_{n = 1}^m {\sum_{i = 1}^{{L}} {|{X_n}[i]{|^2}} }  = \frac{1}{m}\sum_{n = 1}^m {{P_n}}.
\end{align}
Also, independent of the message decoding status, the average transmission power over the packet is ${P^{(M + 1)}} = \frac{1}{{M + 1}}\sum_{n = 1}^{M + 1} {{P_n}} $ if all possible retransmission rounds are used. Therefore, assuming a continuous communication model, the overall long-term average transmission power is determined by
\vspace{-0mm}
\begin{align}
\begin{array}{l}
 \bar P  = \sum_{m = 1}^{M + 1} {{P^{(m)}}\Pr \{ {S_m}\} }  + {P^{(M + 1)}}\Pr \{ {{\hat S}_{M + 1}}\} =\\ \sum_{m = 1}^{M + 1} {(\frac{1}{m}\sum_{n = 1}^m {{P_n}} )\Pr \{ {S_m}\} }  + (\frac{1}{{M + 1}}\sum_{n = 1}^{M + 1} {{P_n}} )\Pr \{ {{\hat S}_{M + 1}}\} \\= \sum_{m = 1}^{M + 1} \bigg({(\frac{1}{m}\sum_{n = 1}^m {{P_n}} )\times} \\\,\,\,\,\,\,\,\Pr \bigg\{ \log (1 + g\sum_{n = 1}^{m - 1} {{P_n}} ) < R
 \le \log (1 + g\sum_{n = 1}^m {{P_n}} )\bigg\}\bigg) \\
  + (\frac{1}{{M + 1}}\sum_{k = 1}^{M + 1} {{P_k}} )\Pr \left\{ \log (1 + g\sum_{n = 1}^{M + 1} {{P_n}} ) < R\right\}  \\
  = \sum_{m = 1}^{M + 1} {(\frac{1}{m}\sum_{n = 1}^m {{P_n}} )\left ({F_G}(\frac{{{e^{{R}}} - 1}}{{\sum_{n = 1}^{m - 1} {{P_n}} }}) - {F_G}(\frac{{{e^{{R}}} - 1}}{{\sum_{n = 1}^m {{P_n}} }})\right )}  \\+ (\frac{1}{{M + 1}}\sum_{k = 1}^{M + 1} {{P_k}} ){F_G}(\frac{{{e^{{R}}} - 1}}{{\sum_{n = 1}^{M + 1} {{P_n}} }}).
 \end{array}
\end{align}

For a given channel gain realization $g$, the data can not be decoded if and only if ${R} > \log (1 + g\sum_{n = 1}^{M + 1} {{P_n}} )$. Therefore, the outage probability constraint $\Pr \{ \text{outage}\}  \le \epsilon$ can be represented as
\vspace{-0mm}
\begin{align}
\begin{array}{l}
 \Pr \left\{ \log (1 + g\sum_{n = 1}^{M + 1} {{P_n}} ) < {R} \right \}  \le \epsilon   \Rightarrow \sum_{n = 1}^{M + 1} {{P_n}}  \ge  \frac{{{e^{{{R}}}} - 1}}{{F_G^{ - 1}(\epsilon )}} \\
 \end{array}
\end{align}
where $F_G^{-1}$ is the inverse function of the channel gain cdf. Intuitively, higher transmission power or more retransmission rounds are needed when the outage probability constraint gets harder, i.e., $\epsilon$ becomes smaller.

%
\vspace{-0mm}
\subsection{Power allocation}
Communication systems may have different power allocation capabilities. For instance, due to, e.g., hardware or complexity limitations, there are cases where, independently of the channel conditions, the data must be transmitted at a fixed power $P$, i.e., $P_m=P\,\forall m$, which is normally called \emph{short-term} power allocation or data transmission with a peak power constraint \cite{5711682,pphdexcelent,isitakhodemun,ekbatanioutage,excellentref,33,5281736,letterkhodemun,5336856,5961851}. In this case, the minimum transmission power satisfying an RTD-based outage probability constraint (10) is found as
\vspace{-0mm}
\begin{align}
P = \frac{{{e^{{{R}}}} - 1}}{{(M + 1)F_G^{ - 1}(\epsilon )}}.
\end{align}

Under the more relaxed long-term (battery-limited) power allocation scenario, the transmitter can adapt the power in each retransmission round. In this case, using (9), the power allocation problem can be stated as \begin{align}
\begin{array}{l}
 \mathop {\min }\limits_{{P_{1},\ldots,{P_{M + 1}}}} \,\,\bar P \\
 \,\,\,\,\,\text{subject}\,\text{to}\,\sum_{n = 1}^{M + 1} {{P_n}}  \ge  \frac{{{e^{{{R}}}} - 1}}{{F_G^{ - 1}( \epsilon )}} \\
 \end{array}
\end{align}
which can be solved numerically.
The following theorem shows that, independent of the number of retransmission rounds, the initial transmission rate and the fading distribution, the transmission powers must be increasing in every RTD-based retransmission round.

\emph{Theorem 1:} In RTD HARQ protocols with an outage probability constraint, the optimal transmission powers, minimizing the average transmission power, must be increasing in every retransmission.
\begin{proof}
The two successive power terms $P_k$ and $P_{k+1}$ in (10) are interchangeable\footnote{As seen in the following, power allocation strategies affect the long-term throughput. However, this point is not important in outage-limited data transmission scenario, as the throughput is not an objective function in this case.}. However, using (9), the contributions of the power terms $P_{k+1}$ and $P_{k}$ on the average transmission power are respectively found as
\vspace{-0mm}
\begin{align}
\begin{array}{l}
 c({P_{k + 1}}) = {P_{k + 1}}{a_{k + 1}}, \\
 {a_{k + 1}} = \sum_{j = k + 1}^{M + 1} {\frac{1}{j} \left ({F_G}(\frac{{{e^R} - 1}}{{\sum_{n = 1}^{j - 1} {{P_n}} }}) - {F_G}(\frac{{{e^R} - 1}}{{\sum_{n = 1}^j {{P_n}} }}) \right )}   \\\,\,\,\,\,\,\,\,\,\,\,\,\,\,\,+ \frac{1}{{M + 1}}{F_G}(\frac{{{e^R} - 1}}{{\sum_{n = 1}^{M + 1} {{P_n}} }})
 \end{array}
\end{align}
and
\vspace{-0mm}
\begin{align}
\begin{array}{l}
 c({P_k}) = {P_k}{a_k}
  \\= {P_k}\bigg ({a_{k + 1}} + \frac{1}{k}\big ({F_G}(\frac{{{e^R} - 1}}{{\sum_{n = 1}^{k - 1} {{P_n}} }}) - {F_G}(\frac{{{e^R} - 1}}{{\sum_{n = 1}^k {{P_n}} }})\big)\bigg ) > {P_k}{a_{k + 1}} \\
 \end{array}
\end{align}
i.e., $a_k \ge a_{k+1}$. Here, $c({P_k})$ is the contribution of the power term $P_k$ on the average transmission power, i.e., $\bar P = \sum_{k = 1}^{M + 1} {c({P_k})}$. Then, using (13) and (14), it is obvious that, in the optimal case, we have $P_k \le P_{k+1}$. This is particularly because with the same powers $P_k=P_{k+1}$ (or $P_k \ge P_{k+1}$) the $k$-th power term has more contribution on the average transmission power than the $(k+1)$-th term. Therefore, in order to have minimum average transmission power, the powers should preferably be given to the last retransmission rounds.
\end{proof}
There is an interesting intuition behind Theorem 1 (see also Fig. 3); implementing RTD HARQ, the initial transmission is set to have a small power. If the channel is \emph{bad} the data can not be decoded and is retransmitted with higher powers. On the other hand, if the channel is \emph{good}, this gambling brings high return. In other words, the HARQ helps us to exploit the good channel conditions properly. Moreover, the outage probability constraint, i.e., (10), does not imply any preference between the order of the power terms. Finally, it is worth noting that in \cite{4200959} it was shown that the same conclusion is valid for basic ARQ schemes where, in each retransmission round, the data is decoded based on the signal received in that time slot, regardless of the previously received signals. However, the conclusion of the theorem is not valid in the problem formulation of \cite{sag5771499}, where the transmission energy is minimized in outage-limited conditions.

\emph{Remark:} Using $\Pr\{{\Upsilon _m}\}=\Pr\{\log(1+g\sum_{n=1}^{m}{P_n})\ge R\}$, $\Pr \{ {{\hat \Upsilon}_1},\ldots,{{\hat \Upsilon}_{m - 1}},{\Upsilon_m}\}=\Pr\{{\Upsilon_m}\}-\Pr\{{\Upsilon_{m-1}}\}$ and some manipulations in (9), it can be written
\begin{align}
\bar P = \sum_{m = 1}^M {(\frac{{\sum_{n = 1}^m {{P_n}}  - m{P_{m + 1}}}}{{m(m + 1)}})\Pr \{ {\Upsilon _m}\} }  + \frac{1}{{M + 1}}{\sum_{m = 1}^{M + 1} {{P_m}} }.\nonumber
\end{align}
According to Theorem 1, we have ${\sum_{n = 1}^m {{P_n}}  - m{P_{m + 1}}}\le 0$. Moreover, using the exponential Chebyshev's inequality, $\Pr \{X \ge x\} \le {e^{ - tx}}{E}({e^{tX}}),\forall t > 0$ \cite{handmath}, with $t=1$, the probability $\Pr\{\Upsilon_m\}$ can be upper bounded by $\Pr \{ {\Upsilon _m}\}  = \Pr \{ \log (1 + g\sum_{n = 1}^m {{P_n}} ) \ge R\}  \le {e^{ - R}}(1 + \lambda \sum_{n = 1}^m {{P_n}} )$ where $\lambda  \buildrel\textstyle.\over= E\{ G\}$. In this way, (10) can be used to bound the average power by
\begin{align}
\bar P &\ge  {e^{ - R}}\sum_{m = 1}^M {\left((\frac{{\sum_{n = 1}^m {{P_n}}  - m{P_{m + 1}}}}{{m(m + 1)}})(1 + \lambda \sum_{n = 1}^m {{P_n}} )\right)} \nonumber\\&+ \frac{{{e^R} - 1}}{{(M + 1)F_G^{ - 1}(\varepsilon )}}.\nonumber
\end{align}
Due to the Chebyshev's inequality, the bound is not tight, particularly at low rates. However, it leads to the following interesting conclusions: 1) the average power scales with $F_G^{-1}(\varepsilon)$ at least inversely linear and 2) with high initial rates $R$, where the first term of the bound vanishes, the outage-limited average power scales with $R$ exponentially. Here, it is interesting to note that, in general, the outage probability is a nonlinear function of the retransmission rates/powers and the average transmission power.


\emph{A simple power allocation algorithm:} Generally, (12) is not a convex optimization problem. Thus, although implementable, gradient-based algorithms are not efficient in this case. To tackle this problem, we propose an iterative algorithm, stated in Algorithm 1, to solve (12). The proposed algorithm has been shown to be efficient in complex optimization problems dealing with local minima issues \cite{springerj,Tcomkhodemun}. However, since the problem is nonconvex, we may need to run the algorithm for several iterations/initial settings, when the number of optimization parameters, i.e., $P_m$'s, increases.

\begin{algorithm} [tbh]
\caption{Power allocation optimization}
\begin{itemize}
\item[I.] For a given outage probability constraint $\Pr \{\text{outage}\}\le \epsilon,$ initial transmission rate $R$ and the fading pdf $f_G$, consider $J$, e.g. $J=20$, randomly generated vectors ${{\Lambda}^j} = [P_1^j,P_2^j,\ldots,P_M^j]$ such that $ P_m^j \ge 0$.
\item[II.] For each vector, do the following procedures
\begin{itemize}
  \item[1)] Determine the last retransmission power $P_{M+1}^j$ according to (10). If $P_{M+1}^j <0$, eliminate the $j$-th vector.
  \item[2)] Determine the average power ${\bar P }^j$ based on (9).
\end{itemize}

\item[III.] Find the vector which results in the lowest average transmission power, i.e., ${{\Lambda}^i}$ where ${ \bar P}^i \le { \bar P}^j,\,\forall j = 1,\ldots,J$.
\item[IV.] ${{\Lambda}^1} \leftarrow {{\Lambda}^i}$.
\item[V.] Generate $b \ll J$, e.g., $b=5$, vectors ${\Lambda}^{j,\text{new}},\,j = 1,\ldots,b$ around ${{{\Lambda}}^1}$. These vectors should also satisfy the constraints introduced in I.
\item[VI.] ${{\Lambda}^{j + 1}} \leftarrow {\Lambda}^{j,\text{new}},\,j = 1,\ldots,b$.
\item[VII.] Regenerate the remaining vectors ${\Lambda}^j,j = b + 2,\ldots,J$ randomly such that ${{\Lambda}^j} = [P_1^j,P_2^j,\ldots,P_M^j]$ and $ P_m^j \ge 0$.
\item[VIII.] Go to II and continue until convergence.
\end{itemize}

\end{algorithm}
\vspace{-4mm}
\subsection{Long-term throughput}
Provided that the data is decoded at the end of the $m$-th (re)transmission round, the transmission rate is $\frac{R}{m}$. Therefore, using (7), the system average rate (or equivalently, throughput) is found as
\vspace{-0mm}
\begin{align}
\begin{array}{l}
 \eta = \sum_{m = 1}^{M + 1} {\frac{R}{m}\Pr \{ S_m\} }   \\\,\,\,\,\,= \sum_{m = 1}^{M + 1} {\frac{R}{m}\left ({F_G}(\frac{{{e^{{R}}} - 1}}{{\sum_{n = 1}^{m - 1} {{P_n}} }}) - {F_G}(\frac{{{e^{{R}}} - 1}}{{\sum_{n = 1}^m {{P_n}} }})\right )}.  \\
 \end{array}
\end{align}
The transmission parameters are determined based on (12). Finally, implementing short-term power allocation, i.e., $P_m=P\,\forall m$, (15) is simplified to
\vspace{-0mm}
\begin{align}
\eta= \sum_{m = 1}^{M + 1} {\frac{R}{m}\left ({F_G}(\frac{{{e^{{R}}} - 1}}{{(m - 1)P}}) - {F_G}(\frac{{{e^{{R}}} - 1}}{{mP}})\right )}.
\end{align}
\vspace{-0mm}
\section{INR HARQ protocols with an outage probability constraint}
Assuming the continuous communication model, this section investigates the performance of the INR HARQ protocol in the presence of an outage probability constraint. Performance analysis for the bursting communication model is presented in the Section VI.

Considering a maximum of $M+1$ INR-based HARQ rounds, $Q$ information nats are encoded into a \emph{mother} codeword of length ${l^{(M + 1)}}$. Then, the codeword is punctured into $M+1$ codewords with powers ${P_m}$ and strictly decreasing rates
\vspace{-0mm}
\begin{equation}
\vspace{-0mm}
{R^{(m)}} = \frac{Q}{{\sum_{j = 1}^m {{l_j}} }} ,\,m = 1,\ldots,M + 1.
\vspace{-0mm}
\end{equation}
Here, ${l_m}$ and $R^{(m)}$ are the channel uses and the \emph{equivalent} transmission rate in the \emph{m}-th time slot, respectively. Moreover, ${l^{(m)}} = \sum_{k = 1}^m {{l_k}}$ denotes the total number of channel uses at the end of the \emph{m}-th slot. Also, ${\xi _n} = {l_n}{P_n}$ is the energy of the signal transmitted at the $m$-th (re)transmission round and ${\xi ^{(m)}} = \sum_{n = 1}^m {{l_n}{P_n}}$ denotes the sum energy consumed in the first $m$ time slots.

The (re)transmission stops at the end of the $m$-th retransmission round if the data is successfully decoded at the $m$-th round and not before. Therefore, implementing random coding and typical set-based decoding, the results of \cite[chapter 15]{excellentref,4444444444} can be used where $\Pr\{S_m\}$ is simplified to the time division multiple access (TDMA)-type equation
\vspace{-0mm}
\begin{align}
\begin{array}{l}
 \Pr \{ {S_m}\}  = \Pr \bigg \{ {R^{(m)}} \le \frac{{\sum_{n = 1}^m {{l_n}\log (1 + g{P_n})} }}{{\sum_{j = 1}^m {{l_j}} }}\,\,{{\cap   }}\\\,\,\,\,\,\,\,\,\,\,\,\,\,\,\,\,\,\,\,\,\,\,\,\,\,\,\,\,\,\,\,\,\,\,\,\,\,\,\,\,\,\,\,\,\,\,\,\,\,\,{R^{(m - 1)}} > \frac{{\sum_{n = 1}^{m - 1} {{l_n}\log (1 + g{P_n})} }}{{\sum_{j = 1}^{m - 1} {{l_j}} }}\,\bigg \}  \\
 \Pr \{ {{\hat S}_{M + 1}}\}  = \Pr \left \{ {R^{(M + 1)}} > \frac{{\sum_{n = 1}^{M + 1} {{l_n}\log (1 + g{P_n})} }}{{\sum_{j = 1}^{M + 1} {{l_j}} }}\right \}.  \\
 \end{array}
\end{align}
Note that, based on (17), we have
\vspace{-0mm}
\begin{align}
\vspace{-0mm}
\frac{{{l_n}}}{{\sum_{j = 1}^m {{l_j}} }} &= {R^{(m)}}\left(\frac{1}{{{R^{(n)}}}} - \frac{1}{{{R^{(n - 1)}}}}\right){\text{ }},\, {R^{(0)}} \buildrel\textstyle.\over= \infty
\vspace{-0mm}
\end{align}
and so $\Pr\{S_m\}$ is found as a function of ${R^{(n)}}$'s.

If the data transmission stops at the end of the $m$-th time slot, the equivalent transmission power is
\vspace{-0mm}
\begin{align}
{P^{(m)}} &= \frac{{{\xi ^{(m)}}}}{{\sum_{j = 1}^m {{l_j}} }} = \frac{{\sum_{n = 1}^m {{l_n}{P_n}} }}{{\sum_{j = 1}^m {{l_j}} }} \nonumber\\&= {R^{(m)}}\sum_{n = 1}^m {{P_n}\left(\frac{1}{{{R^{(n)}}}} - \frac{1}{{{R^{(n - 1)}}}}\right)}
\end{align}
where the last equality is based on (19). Consequently, using (18) and (20), the average transmission power and the average transmission energy per packet are respectively found as
\vspace{-0mm}
\begin{align}
\bar P = \sum_{m = 1}^{M + 1} {{P^{(m)}}\Pr \{ {S_m}\} }  + {P^{(M + 1)}}\Pr \{ {{\hat S}_{M + 1}}\}
\end{align}
and
\vspace{-0mm}
\begin{align}
\bar \xi  = \sum_{m = 1}^{M + 1} {{\xi ^{(m)}}\Pr \{ {S_m}\} }  + {\xi ^{(M + 1)}}\Pr \{ {{\hat S}_{M + 1}}\}.
\end{align}
\subsection{Power allocation}
An outage probability constraint $\Pr \{\text{outage} \} \le \epsilon$ implies that the rate ${R^{(M + 1)}} = \frac{Q}{{{l^{(M + 1)}}}}$ should be, with probability $1-\epsilon$, supported in the last (re)transmission round, i.e.,
\begin{align}
&\Pr \left\{ \frac{{\sum_{n = 1}^{M + 1} {{l_n}\log (1 + g{P_n})} }}{{\sum_{j = 1}^{M + 1} {{l_j}} }} < {R^{(M+1)}}\right\}  = \nonumber\\&\Pr \left\{ \sum_{n = 1}^{M + 1} {\left(\frac{1}{{{R^{(n)}}}} - \frac{1}{{{R^{(n - 1)}}}}\right)\log (1 + g{P_n})}  < {1}\right\}  \le \epsilon.
\end{align}

Therefore, the INR-based outage-limited power allocation problem can be formulated as
\begin{align}
\begin{array}{l}
 \mathop {\min }\limits_{{P_{1}},\ldots,{P_{M + 1}}} \left \{ \sum_{m = 1}^{M + 1} {{P^{(m)}}\Pr \{ {S_m}\} }  + {P^{(M + 1)}}\Pr \{ {{\hat S}_{M + 1}}\} \right \}  \\
 \text{subject}\,\text{to}\,\,\Pr \left \{ \frac{{\sum_{n = 1}^{M + 1} {{l_n}\log (1 + g{P_n})} }}{{\sum_{j = 1}^{M + 1} {{l_j}} }} < {R^{(M + 1)}}\right \}  \le \epsilon.  \\
 \end{array}
\end{align}


The following theorem discusses the optimal adaptive power allocation in the INR protocol with an outage probability constraint.

\emph{Theorem 2: } For INR HARQ protocols with an outage probability constraint, the transmission energies must be increasing in every retransmission.
\begin{proof}
The contributions of the $k$-th and the $(k+1)$-th retransmission rounds on the outage probability constraint (23) are $\frac{{{l_k}\log (1 + g{P_k})}}{{\sum_{j = 1}^{M + 1} {{l_j}} }}$ and $\frac{{{l_{k + 1}}\log (1 + g{P_{k + 1}})}}{{\sum_{j = 1}^{M + 1} {{l_j}} }}$, respectively. Thus, the parameters $(l_m,P_m)$ and $(l_{m+1},P_{m+1})$ are interchangeable, in the sense that the outage probability stays the same if the parameters are switched.

However, this is not the case if we instead study the contributions to the average power. The contributions of the $(k+1)$-th and the $k$-th retransmission rounds on the average transmission power are
\vspace{-0mm}
\begin{align}
\begin{array}{l}
 c({\xi _{k + 1}}) = {\xi _{k + 1}}{b_{k + 1}} = {P_{k + 1}}{l_{k + 1}}{b_{k + 1}}, \\
 {b_{k + 1}} = \sum_{n = k + 1}^{M + 1} {\frac{1}{{\sum_{j = 1}^n {{l_j}} }}\Pr \{ {S_n}\} }  + \frac{1}{{\sum_{j = 1}^{M + 1} {{l_j}} }}\Pr \{ {{\hat S}_1},\ldots,{{\hat S}_{M + 1}}\}  \\
 \end{array}
\end{align}
and
\vspace{-0mm}
\begin{align}
c({\xi _k}) &= {\xi _k}{b_k} = {P_k}{l_k}{b_k} \nonumber\\&= {P_k}{l_k} \left({b_{k + 1}} + \frac{1}{{\sum_{j = 1}^k {{l_j}} }}\Pr \{ {S_k}\}  \right) \ge {\xi _k}{b_{k + 1}},
\end{align}
respectively. Here, $c({\xi_k})$ is the contribution of the energy term $\xi_k$ on the average transmission power, that is, $\bar P = \sum_{k = 1}^{M + 1} {c({\xi_k})}$. Therefore, from (25) and (26), it is obvious that in order to minimize the average transmission power subject to an outage probability constraint, the transmission powers and code lengths should be designed such that the  transmission energies are increasing with the number of retransmissions, i.e., $\xi_k \le \xi_{k+1},\, \forall k$. (For simulations, the readers are referred to Fig. 7.)
\end{proof}

\emph{Remark:} The difference between the performance of different stop-and-wait HARQ protocols is in the way the probability terms $\Pr\{S_m\},\,m=1,\ldots,M+1,$ are calculated, e.g., \cite{5961851} (Also, please see RTD and INR schemes as examples.). Moreover, the proofs of Theorems 1 and 2 are valid independent of how the probabilities $\Pr\{S_m\}$ are determined. Thus, the same conclusion can also be proved for the other stop-and-wait HARQ protocols such as basic ARQ for which the validity of the argument has been previously shown in \cite{4200959}.

As stated before, an outage probability constraint can be satisfied with either a few high-power retransmissions or by many low-SNR retransmission rounds. The following theorem demonstrates the optimal power allocation rule for the second case.

\emph{Theorem 3:} Assume fixed-length coding for the INR. For sufficiently large number of retransmissions, the optimal outage-limited (re)transmission powers follow a specific sequence of numbers which converges to a geometric sequence.
\begin{proof}
With fixed-length coding for the INR, i.e., $l_i=L,\forall i,$ (17) and (20) lead to $R^{(m)}=\frac{R}{m}$ and $P^{(m)}=\frac{1}{m}\sum_{n=1}^m{P_n}$ where $R=\frac{Q}{L}$ is the initial codeword rate. Therefore, for large $M$, (18) and (21) can be used to rephrase the average transmission power (21) as
\begin{align}
\bar P &= R\sum_{m = 1}^{M + 1} {\frac{1}{{m{Z^{(m)}}}}\left({F_G}({Z^{(m - 1)}}) - {F_G}({Z^{(m)}})\right)}  \nonumber\\&+ \frac{R}{{(M + 1){Z^{(M + 1)}}}}{F_G}({Z^{(M + 1)}}).
\end{align}
Here, (27) is obtained by 1) defining ${Z^{(m)}} \buildrel\textstyle.\over= \frac{R}{{\sum_{n = 1}^m {{P_n}} }}$, 2) the fact that the outage probability constraint can be satisfied with low transmission powers when the number of retransmission rounds increases and 3) $\log (1 + x) \to x$ for small values of $x$. The outage probability constraint determines $Z^{(M+1)}$ as in (23). Taking the derivative with respect to $Z^{(m)}$, however, results in
\begin{align}
\frac{{\partial \bar P}}{{\partial {Z^{(m)}}}} &= R(\,\frac{{{F_G}({Z^{(m)}}) - {F_G}({Z^{(m - 1)}})\,}}{{m{Z^{(m)}}^2}}\, \nonumber\\&- \frac{{{f_G}({Z^{(m)}})}}{{m{Z^{(m)}}}} + \frac{{{f_G}({Z^{(m)}})}}{{(m + 1){Z^{(m + 1)}}}}\,)\nonumber\\&\mathop  = \limits^{(a)} Rf_G(Z^{(m)})(\frac{1}{{(m + 1){Z^{(m + 1)}}}} - \frac{{{Z^{(m - 1)}}}}{{m{Z^{(m)}}^2}})\nonumber
\end{align}
where $(a)$ follows from $({F_G}({Z^{(m)}}) - {F_G}({Z^{(m - 1)}})) \to {f_G}({Z^{(m)}})({Z^{(m)}} - {Z^{(m - 1)}})$ for large values of $M$. Therefore, setting $\frac{{\partial \bar P}}{{\partial {Z^{(m)}}}}=0$, the optimal power allocation rule is found by the sequence of numbers
\begin{align}
{Z^{(m)}} = \sqrt {\frac{{m + 1}}{m}{Z^{(m - 1)}}{Z^{(m + 1)}}}.
\end{align}
Particularly, (28) converges to a geometric sequence since ${Z^{(m)}}\to\sqrt{Z^{(m-1)}Z^{(m+1)}}$ when $m$ increases. Finally, it is interesting to note that the power allocation rule (28) is independent of the channel pdf, initial transmission rate and outage probability constraint.
\end{proof}
\emph{Remark}: Using (7), (18) and $\log (1 + x) \to x$ for small values of $x$, we can show that the performance of the RTD and fixed-length coding INR protocols converges when the outage probability constraint gets relaxed or the number of retransmission rounds increases. Thus, the results of Theorem 3 are valid for the RTD as well.

Note that the conclusion of Theorems 2 and 3, which solve the optimal energies minimizing the average power, are different from the results in \cite{ARQGlarsson}, where the expected energy is the optimization criterion.




Finally, considering short-term power allocation, i.e., $P_m=P\,\, \forall m$, the minimum power satisfying the outage probability constraint (23) is obtained by
\vspace{-0mm}
\begin{align}
\Pr \left \{ \log (1 + gP) < {R^{(M+1)}}\right \}  \le \epsilon  \Rightarrow P = \frac{{{e^{{R^{(M+1)}}}} - 1}}{{F_G^{ - 1}(\epsilon )}}.\nonumber
\end{align}

\subsection{Long-term throughput}
Given that the data (re)transmission successfully ends at the end of the $m$-th time slot, the rate $R^{(m)}$ is received at the receiver. Therefore, we can use (18) to find the system throughput as
\vspace{-0mm}
\begin{align}
\begin{array}{l}
 \eta = \sum_{m = 1}^{M + 1} {{R^{(m)}}\Pr \{ {S_m}\} }  \\
  = \sum_{m = 1}^{M + 1} {{R^{(m)}}\Pr \bigg \{ \sum_{n = 1}^{m - 1} {(\frac{1}{{{R^{(n)}}}} - \frac{1}{{{R^{(n - 1)}}}})\log (1 + g{P_n})}  < }\\\,\,\,\,\,\,\,\,\,\,\,\,\,\,\,\,\,\,\,\,\,\,\,\,\,\,\,\,\,\,\,\,\,\,\,\,\,\,\,\,\,\,\,\,\,\,\, 1 \le \sum_{n = 1}^m {(\frac{1}{{{R^{(n)}}}} - \frac{1}{{{R^{(n - 1)}}}})\log (1 + g{P_n})} \bigg \}.  \\
 \end{array}
\end{align}
Also, assuming short-term power constraint (29) is simplified to
\begin{align}
\eta &= \sum_{m = 1}^{M + 1} {{R^{(m)}}\Pr \left\{ {R^{(m)}} \le \log (1 + gP) < {R^{(m - 1)}}\right \} }  \nonumber\\&= \sum_{m = 1}^{M + 1} {{R^{(m)}}\bigg ({F_G}\big(\frac{{{e^{{R^{(m - 1)}}}} - 1}}{P}\big) - {F_G}\big(\frac{{{e^{{R^{(m)}}}} - 1}}{P}\big)\bigg )}.
\end{align}


With the same arguments as in \cite{5961851,5336856}, it can be shown that, compared to the RTD scheme, better performance is achieved in the INR HARQ approach.
Moreover, we can prove that the data is decoded with less (or equal) number of INR-based retransmission rounds, when compared with the RTD. Hence, the expected feedback load and the expected number of retransmission rounds, which are of interest in limited-feedback and delay-sensitive systems, respectively, are less in the INR. However, the superiority of the INR scheme over the RTD is at the cost of higher complexity in both the transmitter and the receiver. This is because the INR HARQ requires variable-length coding at the transmitter, producing new parity bits in each retransmission round and proper combination of the received signals at the receiver.
\vspace{-0mm}
\section{Performance analysis for the bursting communication model}
In this section, the optimal power allocation problem is studied for the bursting communication model, where there is a long idle period between two successive packet transmissions.
\vspace{-0mm}
\subsection{RTD HARQ protocol}

Utilizing the RTD HARQ scheme, the consumed energy is ${\xi ^{(m)}} = L\sum_{n = 1}^m {{P_n}}$, if the data (re)transmission ends at the end of the $m$-th (re)transmission round. Therefore, the expected energy within a packet transmission period is obtained by
\vspace{-0mm}
\begin{align}
E\{ \tilde \xi \} & = L\sum_{m = 1}^{M + 1} {\left(\sum_{n = 1}^m {{P_n}} \right)\Pr \{ {S_m}\} }  + \left(L\sum_{n = 1}^{M + 1} {{P_n}} \right)\Pr \{ {{\hat S}_{M + 1}}\} \nonumber\\&\mathop  = \limits^{(b)} L\left ({P_1} + \sum_{m = 2}^{M + 1} {{P_m}{F_G}(\frac{{{e^R} - 1}}{{\sum_{n = 1}^{m - 1} {{P_n}} }})} \right)
\end{align}
where $(b)$ follows from (7) and some straightforward calculations. On the other hand, $mL$ channel uses are spent in the first $m$ rounds. Also, independent of the message decoding status, there will be $(M+1)L$ channel uses if all possible retransmission rounds are used. Therefore, the expected number of channel uses and the average transmission power are respectively found as
\vspace{-0mm}
\begin{align}
E\{ \tilde \tau \} & = L\sum_{m = 1}^{M + 1} {m\Pr \{ {S_m}\} }  + (M + 1)L\Pr \{ {{\hat S}_{M + 1}}\}\nonumber\\& \mathop  = \limits^{(c)} L\left (1 + \sum_{m = 1}^M {{F_G}(\frac{{{e^R} - 1}}{{\sum_{n = 1}^m {{P_n}} }})} \right)
\end{align}
and
\vspace{-0mm}
\begin{align}
\varphi  = \frac{{E\{ \tilde \xi \} }}{{E\{ \tilde \tau \} }} = \frac{{{P_1} + \sum_{m = 2}^{M + 1} {{P_m}{F_G}(\frac{{{e^R} - 1}}{{\sum_{n = 1}^{m - 1} {{P_n}} }})} }}{{1 + \sum_{m = 1}^M {{F_G}(\frac{{{e^R} - 1}}{{\sum_{n = 1}^m {{P_n}} }})} }}
\end{align}
where $(c)$ follows from (7). In this way, using (10), the power allocation problem can be formulated as
\vspace{-0mm}
\begin{align}
\begin{array}{l}
 \mathop {\min }\limits_{{P_1},\ldots,{P_{M + 1}}} \{ \frac{{{P_1} + \sum_{m = 2}^{M + 1} {{P_m}{F_G}(\frac{{{e^R} - 1}}{{\sum_{n = 1}^{m - 1} {{P_n}} }})} }}{{1 + \sum_{m = 1}^M {{F_G}(\frac{{{e^R} - 1}}{{\sum_{n = 1}^m {{P_n}} }})} }}\}  \\
 \text{subject}\,\text{to}\,\sum_{m = 1}^{M + 1} {{P_m}}  \ge \frac{{{e^R} - 1}}{{F_G^{ - 1}(\epsilon )}} \\
 \end{array}
\end{align}
which can be solved numerically (Simulation results can be found in Figs. 4-6).
\subsubsection{Long-term throughput}
Provided that the receiver can decode the data, $Q$ nats are received by the receiver in each packet. Therefore, the expected received information nats in each packet is
\vspace{-0mm}
\begin{align}
E\{\tilde Q\} & = {Q}\left (1 - \Pr \left\{ R \ge \log (1 + g\sum_{n = 1}^{M + 1} {{P_n}} )\right\} \right )\nonumber\\&={Q\left (1 - {F_G}(\frac{{{e^{{R}}} - 1}}{{\sum_{n = 1}^{M + 1} {{P_n}} }})\right )}.
\end{align}
Consequently, using (32) and as $R=\frac{Q}{L}$, the system throughput is obtained by
%
\vspace{-0mm}
\begin{align}
\eta  = \frac{{R(1 - {F_G}(\frac{{{e^R} - 1}}{{\sum_{n = 1}^{M + 1} {{P_n}} }}))}}{1 + \sum_{m = 1}^M {{F_G}(\frac{{{e^R} - 1}}{{\sum_{n = 1}^m {{P_n}} }})}}
\end{align}
where the transmission powers are determined by (34). Finally, implementing short-term power allocation, i.e., $P_m=P\,\forall m$, the system throughput is rephrased as
\vspace{-0mm}
\begin{align}
\eta  = \frac{{R\big(1 - {F_G}(\frac{{{e^R} - 1}}{{(M + 1)P}})\big)}}{1 + \sum_{m = 1}^M {{F_G}(\frac{{{e^R} - 1}}{{mP}})} }.
\end{align}
\vspace{-0mm}
\subsection{INR HARQ protocol}
Using (18) and (22), the expected energy consumed within an INR-based packet transmission period is found as
\vspace{-0mm}
\begin{align}
E\{ \tilde \xi \}  = {P_1}{l_1} + \sum_{n = 2}^{M + 1} {{P_n}{l_n}\Pr \left\{ \sum_{j = 1}^{n - 1} {\frac{{{l_j}\log (1 + g{P_j})}}{{\sum_{i = 1}^{n - 1} {{l_i}} }} < {R^{(n - 1)}}} \right\} }.
\end{align}

Also, with the same arguments as for (32), the expected number of channel uses in each packet is
\vspace{-0mm}
\begin{align}
E\{ \tilde \tau \}&  = \sum_{m = 1}^{M + 1} {(\sum_{n = 1}^m {{l_n}} )\Pr \{ {S_m}\} }  + (\sum_{n = 1}^{M + 1} {{l_n}} )\Pr \{ {{\hat S}_{M + 1}}\} \nonumber\\&\mathop  = \limits^{(d)} {l_1} + \sum_{n = 2}^{M + 1} {{l_n}\Pr \{ \sum_{j = 1}^{n - 1} {\frac{{{l_j}\log (1 + g{P_j})}}{{\sum_{i = 1}^{n - 1} {{l_i}} }} < {R^{(n - 1)}}} \} }
\end{align}
where $(d)$ comes from (18). Hence, from (17), the average INR-based transmission power is obtained by
\begin{align}
&\varphi  = \frac{{\frac{{{P_1}}}{{{R^{(1)}}}} + \sum_{n = 2}^{M + 1} {{P_n}(\frac{1}{{{R^{(n)}}}} - \frac{1}{{{R^{(n - 1)}}}})\Theta_n } }}{{\frac{1}{{{R^{(1)}}}} + \sum_{n = 2}^{M + 1} {(\frac{1}{{{R^{(n)}}}} - \frac{1}{{{R^{(n - 1)}}}})\Theta_n } }}, \nonumber\\& \Theta_n \buildrel\textstyle.\over=\Pr \{ \sum_{j = 1}^{n - 1} {(\frac{1}{{{R^{(j)}}}} - \frac{1}{{{R^{(j - 1)}}}})\log (1 + g{P_j}) < 1} \}.
\end{align}
Replacing (40) in (24), the power allocation problem can solved numerically (Please see Fig. 6).
\subsubsection{Long-term throughput}
%
%
$Q$ information nats are received if the data is decoded at any retransmission round. Therefore, the INR-based expected received information nats in each packet is
\vspace{-0mm}
\begin{align}
E\{\tilde Q\}  &= Q \big(1 - \Pr \{ {{\hat S}_{M + 1}}\} \big) \nonumber\\&= Q\Pr \big\{ \sum_{n = 1}^{M + 1} {(\frac{1}{{{R^{(n)}}}} - \frac{1}{{{R^{(n - 1)}}}})\log (1 + g{P_n})}  \ge 1\big\}
\end{align}
and, using (17) and (39), the long-term throughput is obtained by
\begin{align}
\eta  = \frac{{E\{ \tilde Q\} }}{{E\{ \tilde \tau \} }} = \frac{{\Pr \left\{ {\sum_{n = 1}^{M + 1} {(\frac{1}{{{R^{(n)}}}} - \frac{1}{{{R^{(n - 1)}}}})\log (1 + g{P_n})}  \ge 1} \right\}}}{{\frac{1}{{{R^{(1)}}}} + \sum_{n = 2}^{M + 1} {(\frac{1}{{{R^{(n)}}}} - \frac{1}{{{R^{(n - 1)}}}})\Theta_n } }}.
\end{align}
Finally, assuming short-term power constraint (42) simplifies to
\vspace{-0mm}
\begin{align}
\eta  = \frac{{1 - {F_G}(\frac{{{e^{{R^{(M + 1)}}}} - 1}}{P})}}{{\sum_{m = 1}^{M + 1} {\frac{{{F_G}(\frac{{{e^{{R^{(m - 1)}}}} - 1}}{P}) - {F_G}(\frac{{{e^{{R^{(m)}}}} - 1}}{P})}}{{{R^{(m)}}}}}  + \frac{{{F_G}(\frac{{{e^{{R^{(M + 1)}}}} - 1}}{P})}}{{{R^{(M + 1)}}}}}}.
\end{align}

Due to the denominators of (33) and (40), it is difficult to prove the validity of Theorems 1, 2 and 3 under bursting communication model, although they seem intuitively valid. However, the following theorem studies the power allocation problem for simplest cases of RTD and INR HARQ feedback where the validity of the previous arguments is proved for the case of $M=1$ bit ARQ feedback.

\emph{Theorem 4:} Considering one bit feedback in the RTD and the INR HARQ protocols under bursting communication model, the following assertions are valid if $R^{(m)}=\frac{R}{m},\,m=1,2,$ where $R^{(m)}$ and $\frac{R}{m}$ are the INR- and RTD-based equivalent (re)transmission rates in the $m$-th round, respectively:
\begin{itemize}
  \item[(a)] In both schemes, the transmission powers should be increasing with the number of retransmission rounds.
  \item[(b)] In order to satisfy an outage probability constraint, less average transmission power is required in the INR HARQ protocol\footnote{As stated before, the same argument as in part (b) can be proved for arbitrary number of retransmissions, if the continuous communication model is considered.}.
\end{itemize}
\begin{proof}
Please see Appendix A.
\end{proof}
Finally, note that if there is an additional constraint on the peak transmission power, the adaptive power controllers can be easily updated by the same procedure as in \cite[Section III.B]{33} which was developed for throughput maximization in communication setups utilizing quantized CSI.
\vspace{-0mm}
\section{Discussions}
In this section, we present some discussions on the system model assumptions and possible extensions of the paper.
\vspace{-0mm}
\subsection{On temporal variations of the fading coefficients}
Block-fading is an appropriate model for the stationary or slow-moving users \cite{5281736,letterkhodemun,ARQGlarsson,sag5771499,excellentref,5336856}. For the users with high speed or long codewords, compared to the channel coherence time, one can consider the fast-fading models where each codeword spans a fading block, and the channel changes in each (re)transmission round \cite{5961851}\footnote{Under fast-fading channel conditions, the INR protocol is studied with fixed-length coding because the length of the codewords is the same as the fading block length.}. In this case, while the average transmission power, i.e., (9), (21), (33) and (40), is obtained with the same procedure as before, the probability terms $\Pr\{S_m\}$ and $\Pr\{\hat S_{M+1}\}$ are replaced by
\begin{align}
\Pr\{S_m\}=\left\{\begin{matrix}
\Pr\{\log(1+\sum_{n=1}^{m-1}{g_nP_n})<R\\\,\,\,\,\,\,\,\,\,\,\,\,\,\,\,\,\,\,\,\,\,\,\,\,\,\le \log(1+\sum_{n=1}^{m}{g_nP_n})\}, & \text{For RTD}\\
\Pr\{\sum_{n=1}^{m-1}{\log(1+g_nP_n)}<R\\\,\,\,\,\,\,\,\,\,\,\,\,\,\,\,\,\,\,\,\,\,\,\,\,\,\le \sum_{n=1}^{m}{\log(1+g_nP_n)}\}, & \text{For INR}
\end{matrix}\right.\nonumber
\end{align}
\begin{align}
\Pr\{\hat S_{M+1}\}=\left\{\begin{matrix}
\Pr\{\log(1+\sum_{n=1}^{M+1}{g_nP_n})<R\} & \text{For RTD}\\
\Pr\{\sum_{n=1}^{M+1}{\log(1+g_nP_n)}<R\} & \text{For INR}
\end{matrix}\right.\nonumber
\end{align}
where $g_m$ is the channel realization at the $m$-th round. Here, the interesting point is that the general conclusions of the paper, e.g., Theorems 1, 2 and 4, hold for the fast-fading scenario as well, since the arguments are valid for every given probability terms $\Pr\{S_m\}$ and $\Pr\{\hat S_{M+1}\}$, independent of how they are found (For simulation results with fast-fading conditions, please see Figs. 12-14.).

Finally, with block-fading conditions and, e.g., a continuous communication model, it may occur that two different channel realizations are experienced during the last packet transmission of a fading block. In this case, the results of the fast- and block-fading models can be combined to study the system performance during the transmission of that packet. However, with sufficiently long fading blocks, we can ignore the effect of that single packet, as in \cite{5281736,letterkhodemun,ARQGlarsson,sag5771499,excellentref,5336856}.
\vspace{-0mm}
\subsection{On the effect of the feedback channel}
Throughout the paper, the feedback signal is supposed to be error-free. This is an acceptable assumption in many scenarios because the feedback bits are sent at very low rates, compared to the forward link, and with an appropriate coding they can be decoded error-free \cite{letterkhodemun,ARQGlarsson,sag5771499,excellentref,5336856,cairearq1,33}. Also, the energy consumption in the feedback link is normally much less than the one in the forward link, such that it is ignored in the optimization problem \cite{ARQGlarsson,sag5771499,isitakhodemun,excellentref,5711682,ekbatanioutage,outageHARQ}. A more accurate extension of the paper, however, is to take the power consumption in the feedback channel into account. In that case, there is a tradeoff between the feedback channel reliability and the power consumption in the feedback/forward links; with a low-power feedback signal the feedback bits may be decoded erroneously and, consequently, lead to waste of energy/throughput resources in the forward/feedback links. Thus, depending on the feedback channel fading model, power adaptation capability and the coding scheme of the feedback bits, a wide range of results can be observed in the cases with an imperfect feedback channel.
\vspace{-0mm}
\subsection{Performance improvement via reinforcement algorithms}
In harmony with practical HARQ protocols \cite{5961851,pphdexcelent,letterkhodemun,5336856,cairearq1,cairearq2,ARQ20112,ARQ20113,sag5771499,ARQGlarsson,1200407,Arulselvan,4356994,5497972,4200959}, we considered the same set of (re)transmission rates and powers for each packet transmission. However, for the continuous communication model, the system data transmission efficiency can be improved via implementation of a reinforcement algorithm \cite{reinforcement}\footnote{Reinforcement algorithms refer to the schemes where the results of the previous trials and a reward-punishment rule are used for parameter setting in the next steps \cite{reinforcement}. Due to the long idle period between successive packet transmissions, the reinforcement schemes are not useful in the bursting communication model.}. In that setup, not only the (re)transmission rates/powers are different in each round but also the whole parameters are adapted in each packet period based on the decisions in the previous packet transmissions. Thus, the performance is improved, as the system adaptation capability increases (For simulation results, please see Fig. 14). However, it is worth noting that implementation of reinforcement schemes increases the implementation complexity substantially. Also, many practical HARQ schemes are designed to operate at fixed (re)transmission rates/powers, for which reinforcement algorithms are not feasible. Finally, the reinforcement-based schemes suffer from the error propagation problems where erroneous decoding of a single feedback bit, which may occur in some practical scenarios, affects many successive packet transmissions.
\subsection{Sub-packeting approach}
In this paper, we concentrate on a \emph{complete packeting} scheme where $Q$ nats are encoded into a, e.g., RTD-based codeword of length $L$ and the whole codeword is repeated in the retransmissions \cite{1632107}. An alternative scheme is the \emph{sub-packeting} approach where each codeword is constructed by concatenation of $N$ parallel error detection encoders, producing subcodewords of length $\frac{L}{N}$, and one forward error correction encoder \cite[Fig. 2]{1632107}. In this case, only the erroneous subcodewords, and not the whole codeword, are retransmitted in the retransmission rounds which leads to energy saving and outage probability/transmission delay reduction. The sub-packeting scheme is of interest in the scenario of bursty errors and when short length cyclic redundancy check (CRC) codes are used, and is more flexible than the complete packeting approach. Therefore, an interesting extension of the paper is to study the effect of adaptive power allocation on the performance of the sub-packeting approach when the channel experiences very-fast-fading conditions, e.g., \cite{1632107}, and there is burst error probability. Here, the same approach as before can be used for power adaptation between the subcodewords.
\vspace{-0mm}
\section{Simulation results}
Simulation results are given for Rayleigh-fading channels $f_G(g)=\lambda e^{-\lambda g},\,g \ge 0$ where we set $\lambda =1.$ Also, the initial transmission rate is set to $R=1$, unless otherwise stated. We consider fixed-length coding for the INR, i.e., $R^{(m)}=\frac{R}{m}$. Further, in all figures, the powers are presented in dB. In Figs. 3-11, we consider the block-fading condition; For the bursting model, a different gain realization is experienced in each packet period. With a continuous communication model, many packets are sent in each fading block. In Figs. 12-14, we study the system performance under fast-fading assumption where the channel changes in each retransmission round.

\begin{figure}
\vspace{-0mm}
\centering
  \includegraphics[width=0.98\columnwidth]{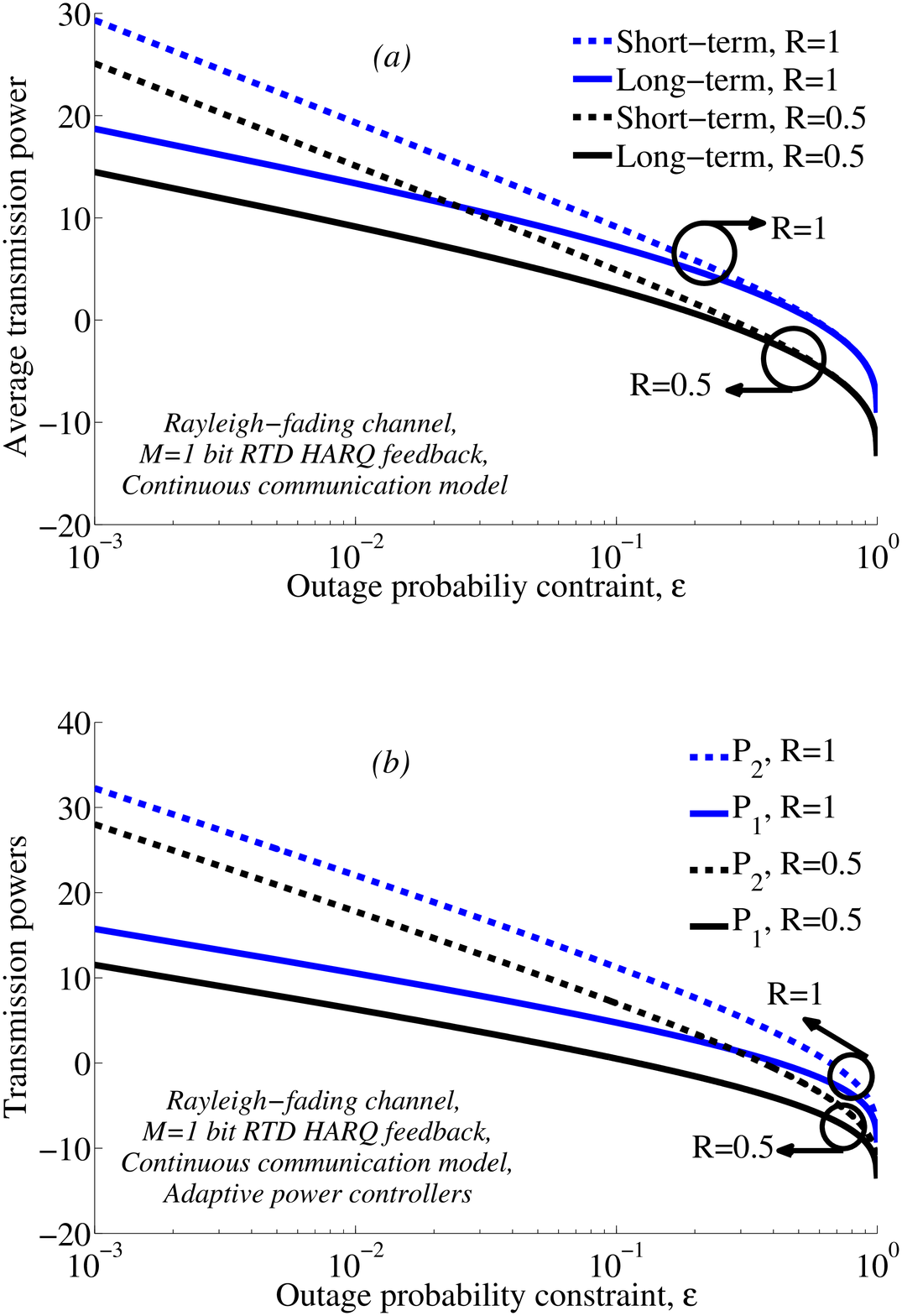}\\\vspace{-6mm}
\caption{(a) Average transmission power and (b) transmission powers in the (re)transmission rounds as a function of outage probability constraint, Rayleigh-fading channel, $M=1$ bit RTD HARQ protocol $R=1$ or $0.5$, continuous communication. The required transmission power increases with the initial transmission rate $R$. Further, in the optimal case higher power is given to the second retransmission round, compared to the first round.  }\label{figure111}
\vspace{-4mm}
\end{figure}

\begin{figure}
\vspace{-0mm}
\centering
  \includegraphics[width=0.98\columnwidth]{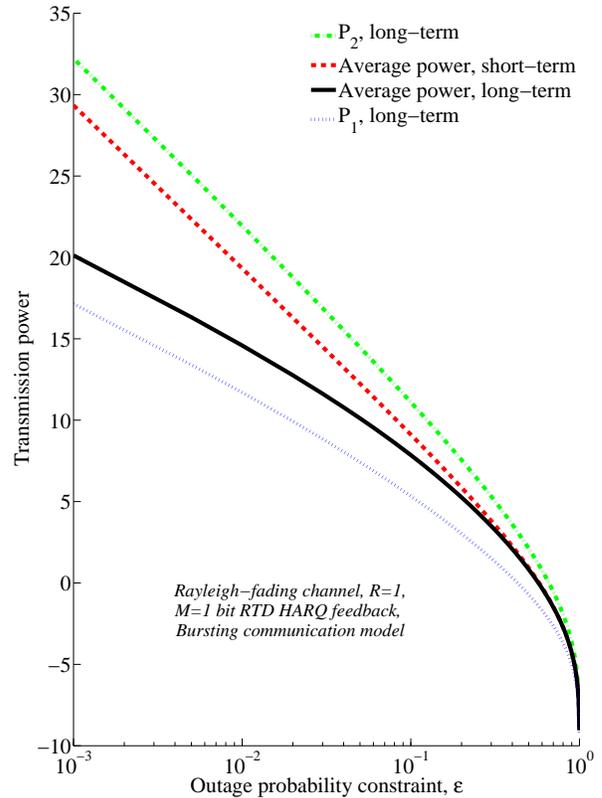}\\\vspace{-8mm}
\caption{Transmission power vs outage probability constraint, Rayleigh-fading channel, $M=1$ bit RTD HARQ feedback $R=1$, different power allocation scenarios. Under long-term power allocation condition, the optimal transmission power in the first round is less than the one in the second round.}\label{figure111}
\vspace{-4mm}
\end{figure}

\begin{figure}
\vspace{-0mm}
\centering
  \includegraphics[width=0.98\columnwidth]{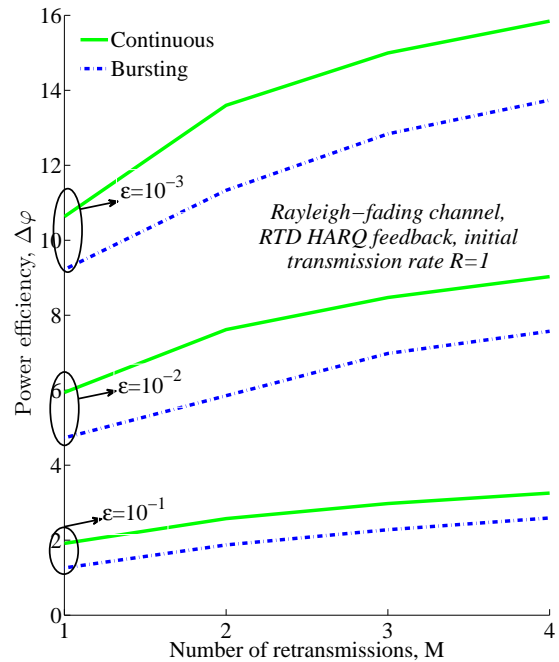}\\\vspace{-4mm}
\caption{Power efficiency $\Delta \varphi=\varphi^\text{short-term}-\varphi^\text{long-term}$ for different number of retransmission rounds $M$. Rayleigh-fading channel, RTD HARQ feedback $R=1$. The data is (re)transmitted a maximum of $(M+1)$ rounds.}\label{figure111}
\vspace{-4mm}
\end{figure}

Utilizing the RTD HARQ protocol, Figs. 3-5 investigate the effect of the outage probability constraint on the required transmission power under continuous and bursting communication models. The figures emphasize the validity of Theorems 1 and 4. Optimal power allocation brings sizable gains; with an outage probability ${10}^{-3}$ and $M=1$, optimal power allocation reduces the average power by 9 and 11 dB in the bursting and continuous communication models, respectively, compared to short-term power allocation (Figs. 3-5). Finally, Fig. 5 studies the power efficiency, which is defined as $\Delta \varphi= \varphi^\text{short-term}-\varphi^\text{long-term}$ with $\varphi^\text{short-term}$ and $\varphi^\text{long-term}$ being the average power achieved by short- and long-term power allocation, respectively. As illustrated in the figures, the effect of long-term power allocation decreases under relaxed outage probability constraints and increases with the number of retransmission rounds.

\begin{figure}
\vspace{-0mm}
\centering
  \includegraphics[width=0.9\columnwidth]{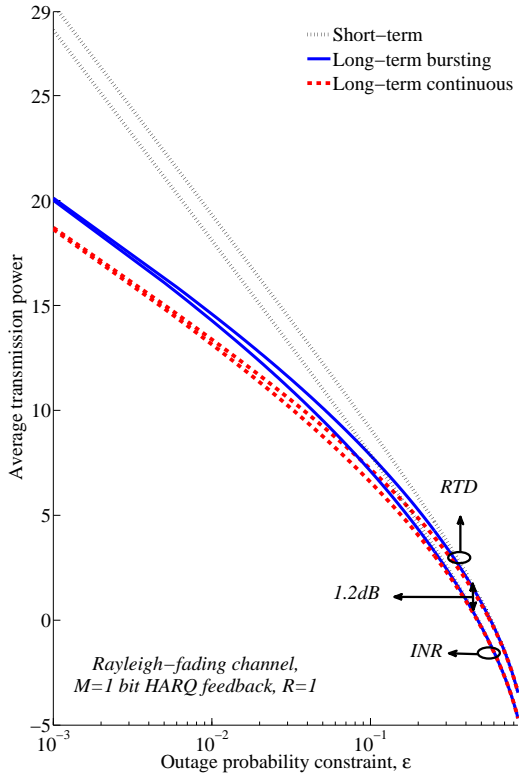}\\\vspace{-4mm}
\caption{Average transmission power vs outage probability constraint, Rayleigh-fading channel, $M=1$ bit HARQ feedback $R=1$, different power allocation scenarios. With high outage probability constraints, the average transmission power of the INR HARQ scheme is $1.2 dB$ less than the one in the RTD protocol. Further, less transmission power is required to satisfy an outage probability constraint in the continuous communication model when compared with the bursting model. }\label{figure111}
\vspace{-4mm}
\end{figure}

\begin{figure}
\vspace{-0mm}
\centering
  \includegraphics[width=0.9\columnwidth]{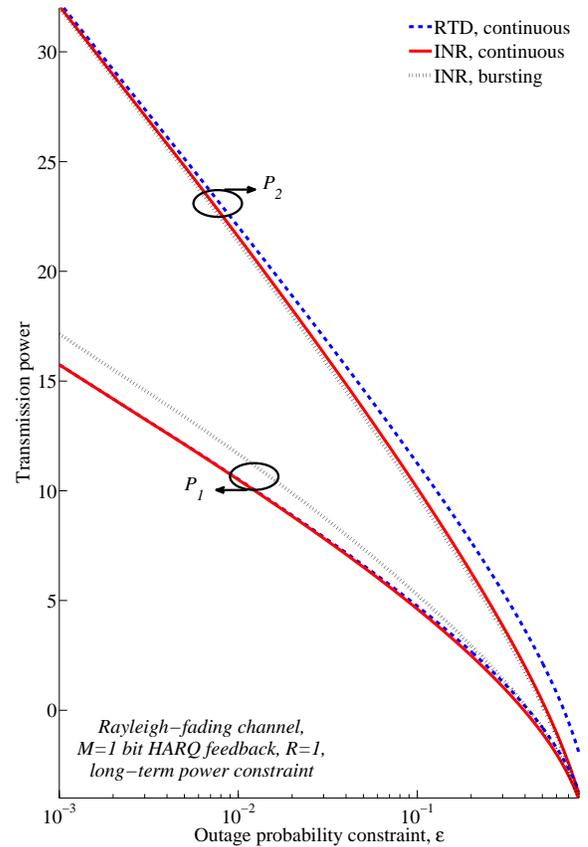}\\\vspace{-4mm}
\caption{The first and the second rounds transmission power in different HARQ schemes and communication models, Rayleigh-fading channel, $M=1$ bit HARQ feedback $R=1$, adaptive power allocation scenario. The effect of power allocation diminishes under relaxed outage probability constraints, i.e., $\epsilon \to 1.$ }\label{figure111}
\vspace{-4mm}
\end{figure}

Figs. 6 and 7 demonstrate the effect of an outage probability constraint on the required transmission power of the INR protocol and compare the results with the ones obtained for the RTD protocol. Here, the results show that 1) with medium outage probability constraints, the average transmission power of the INR HARQ scheme is about $1.2$dB less than the one in the RTD protocol (Fig. 6). 2) Less transmission power is required to satisfy an outage probability constraint in the continuous communication model when compared with the bursting model. However, this difference diminishes as the outage probability constraint becomes more relaxed, i.e., $\epsilon$ becomes larger (Fig. 6). 3) Considerable performance improvement is achieved with adaptive power allocation in hard outage probability constraints, i.e., small $\epsilon$'s. For instance, comparing with short-term power allocation at outage probability ${10}^{-3}$ and $M=1$, adaptive power allocation leads to 8 and 9dB average power reduction in bursting and continuous communication models, respectively (Fig. 6). However, the effect of power allocation diminishes under relaxed outage probability constraints (Figs. 6 and 7). Moreover, Figs. 3, 4 and 7 emphasize that with long-term power allocation and for both protocols, the optimal transmission power in the first round is less than the one in the second round, in harmony with Theorems 1, 2 and 4.

\begin{figure}
\vspace{-0mm}
\centering
  \includegraphics[width=0.9\columnwidth]{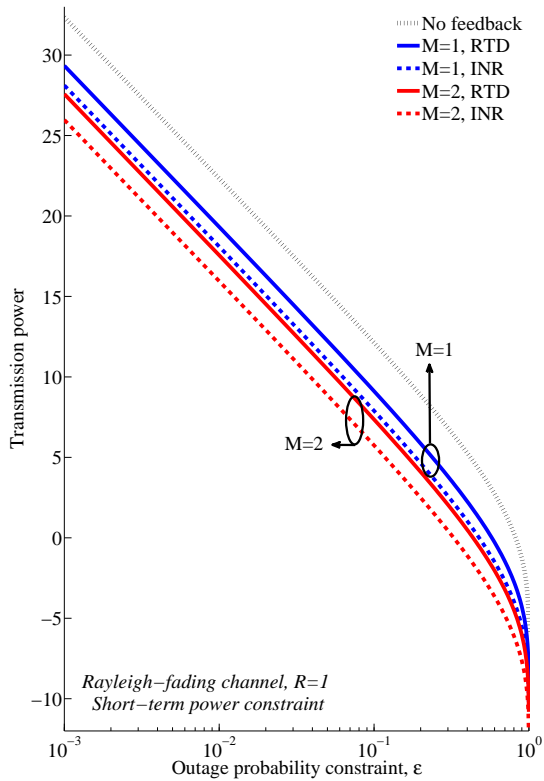}\\\vspace{-4mm}
\caption{Transmission power vs outage probability constraint, short-term power constraint, Rayleigh-fading channel, $M=1\, \text{or}\, 2$ bits HARQ feedback, $R=1$. The required transmission power satisfying an outage probability constraint decreases with the number of retransmission rounds. }\label{figure111}
\vspace{-4mm}
\end{figure}

Considering $M=1$ and 2 retransmission rounds, Fig. 8 shows the transmission power as a function of the outage probability constraint under short-term power allocation condition. In particular, Fig. 8 (Fig. 3) indicates that the required transmission power satisfying an outage probability constraint decreases (increases) with the number of retransmission rounds (the initial transmission rate $R$). Moreover, comparing Figs. 3-8, optimal power allocation is much more effective than increasing the number of retransmission rounds in improving the outage-limited performance of HARQ protocols.
\begin{figure}
\vspace{-0mm}
\centering
  \includegraphics[width=0.9\columnwidth]{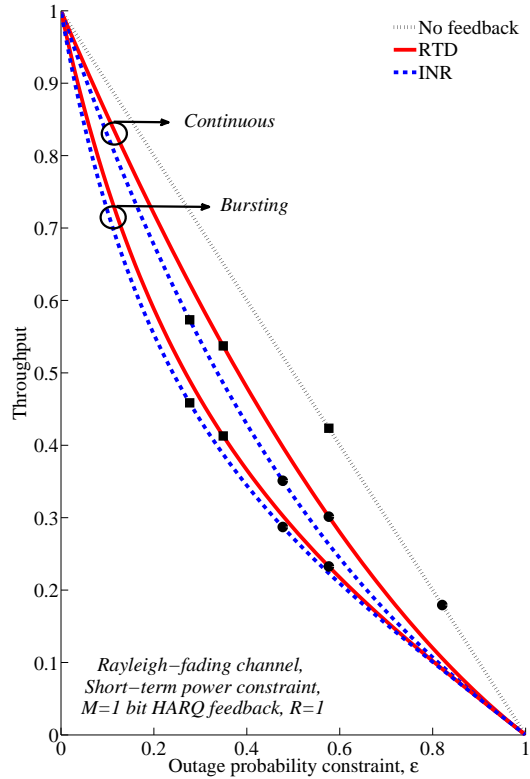}\\\vspace{-6mm}
\caption{Throughput vs outage probability constraint, short-term power constraint, Rayleigh-fading channel, $M=1$ bit HARQ feedback, $R=1$. Circles (squares) represent the results with transmission power $P=1$ ($P=2$). With the same transmission power, less outage probability and higher throughput are achieved in the INR protocol, compared to RTD. Also, higher throughput is achievable in the continuous model, in comparison with the bursting model.  }\label{figure111}
\vspace{-6mm}
\end{figure}
\begin{figure}
\vspace{-0mm}
\centering
  \includegraphics[width=0.93\columnwidth]{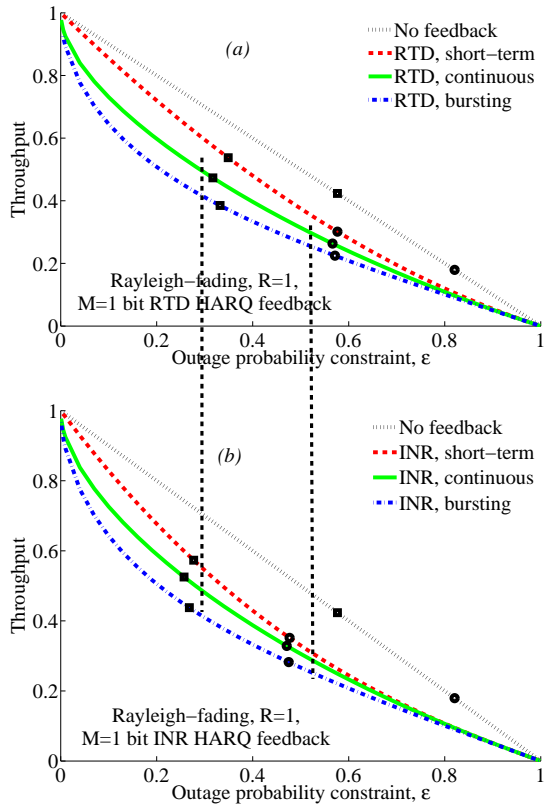}\\\vspace{-4mm}
\caption{Throughput vs outage probability constraint in (a) RTD or (b) INR protocols, different power allocation conditions, Rayleigh-fading channel, $M=1$ bit HARQ feedback, R=1. Circles (squares) represent the results with average transmission power $P=1$ ($P=2$). In all schemes, INR outperforms the RTD in terms of throughput and outage probability, for a given average transmission power. Further, higher throughput and lower outage probability is achieved in the continuous communication model, in comparison to the bursting model. With the same average powers, long-term power allocation wins (loses) in competition with short-term power allocation in terms of outage probability (throughput). }\label{figure111}
\vspace{-4mm}
\end{figure}
\begin{figure}
\vspace{-0mm}
\centering
  \includegraphics[width=0.93\columnwidth]{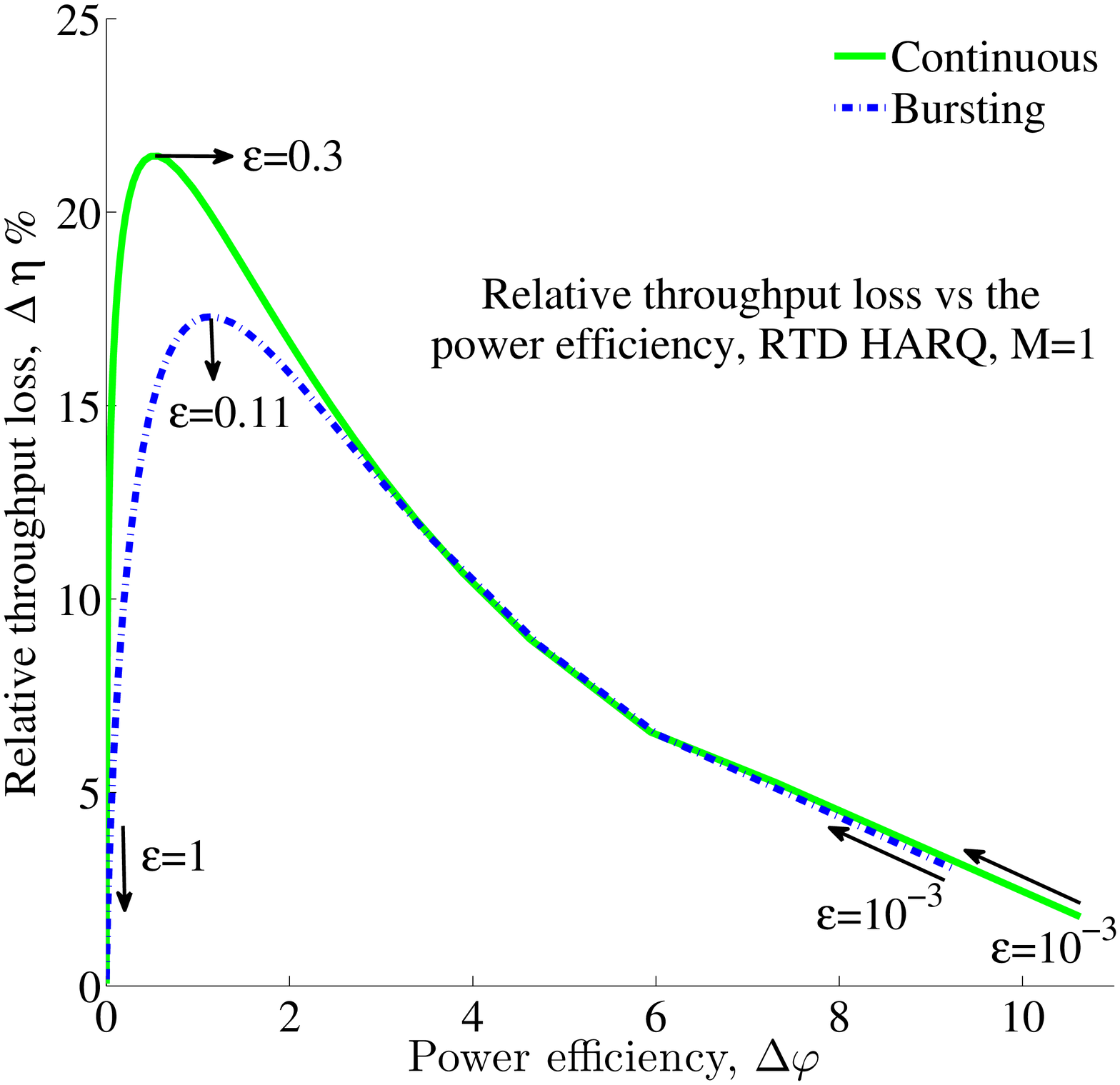}\\
\caption{Relative throughput loss $\Delta\eta=\frac{\eta^\text{short-term}-\eta^\text{long-term}}{\eta^\text{long-term}}\%$ vs the power efficiency $\Delta \varphi$ dB. RTD HARQ protocol, Rayleigh-fading channel, $M=1$ bit feedback, $R=1$. Down right is the desired region in the figure. }\label{figure111}
\vspace{-0mm}
\end{figure}

Figures 9-11 investigate the system throughput versus the outage probability constraint in different conditions. Here, it is emphasized that with the same transmission power, less outage probability and higher throughput are achieved for the INR protocol, compared to RTD. Also, higher throughput and less (or equal) outage probability are achievable in the continuous model, in comparison with bursting model (Fig. 9 and 10). Moreover, considering the same powers, long-term power allocation wins (loses) in competition with short-term power allocation in terms of outage probability (throughput) (Fig. 10). This point is further evaluated in Fig. 11 where the relative throughput loss $\Delta\eta=\frac{\eta^\text{short-term}-\eta^\text{long-term}}{\eta^\text{long-term}}\%$ is plotted versus the power efficiency $\Delta \varphi$. Here, ${\eta^\text{short-term}}$ and ${\eta^\text{long-term}}$ denote the throughput achieved by short- and long-term power allocation, respectively. Interestingly, the relative throughput loss, due to outage-limited power allocation, diminishes at relaxed and hard outage probability constraints, because the required average power goes to zero and infinity, respectively. Also, the figures emphasize that different optimal power allocation strategies are obtained, depending on whether the outage probability or the system throughput is considered as the optimization criterion. That is, the power-limited outage minimization and the power-limited throughput maximization problems are not interchangeable.

As mentioned before, block-fading is a coarse but effective theoretical model of the fading channels, particularly for the stationary/slow-moving users. In Figs. 12-14, we analyze the system performance with the more realistic correlated time-varying fading model of \cite[eq. 12]{6051400} (and the references therein) where the channel changes in each codeword transmission period according to a first-order Gauss-Markov process
\begin{align}
h_{k+1}=\beta h_k+\sqrt{1-\beta^2}\vartheta,\,\vartheta\sim \mathcal{CN}(0,1).\nonumber
\end{align}
Here, $\beta$ is the correlation factor of fading realizations experienced in two successive codeword transmissions with $\beta=0$ (resp. $\beta=1$) representing the uncorrelated fast-fading (resp. block-fading) channel model. With this model, different channel realizations are experienced during a packet period. The simulation results indicate that the general conclusions of the paper, e.g., Theorems 1, 2 and 4, hold independently of the fading channel characteristics (Please see Subsection VII.A as well). For a given outage probability, the average transmission power increases with temporal correlation of the fading channel (Fig. 12). However, the gain of optimal power allocation is almost constant for different correlation conditions (Fig. 13). Finally, for further discussions about the fading channel models, the readers are referred to, e.g., \cite{cooperationlimit,5567190,6093903,720551}, which present remarkable equivalences between the block-fading and continuous fading models, when the Doppler spectrum of the continuous fading model is bandlimited.
\begin{figure}
\vspace{-0mm}
\centering
  \includegraphics[width=0.92\columnwidth]{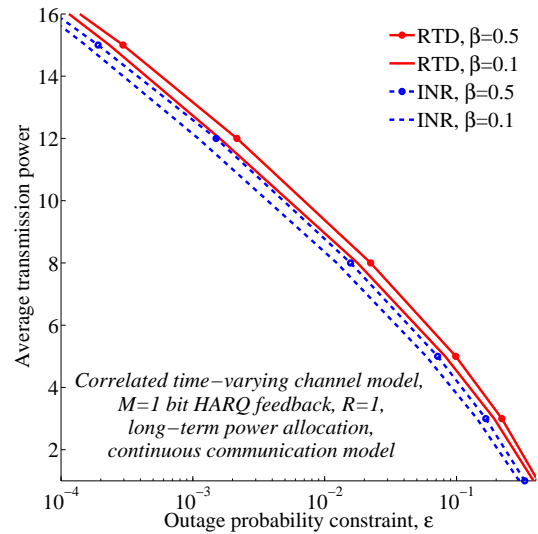}\\
\caption{Performance analysis for the correlated time-varying fading model of \cite{6051400}, $M=1$ bit feedback, $R=1$, long-term power allocation, continuous communication model.  }\label{figure111}
\vspace{-4mm}
\end{figure}
\begin{figure}
\vspace{-0mm}
\centering
  \includegraphics[width=0.92\columnwidth]{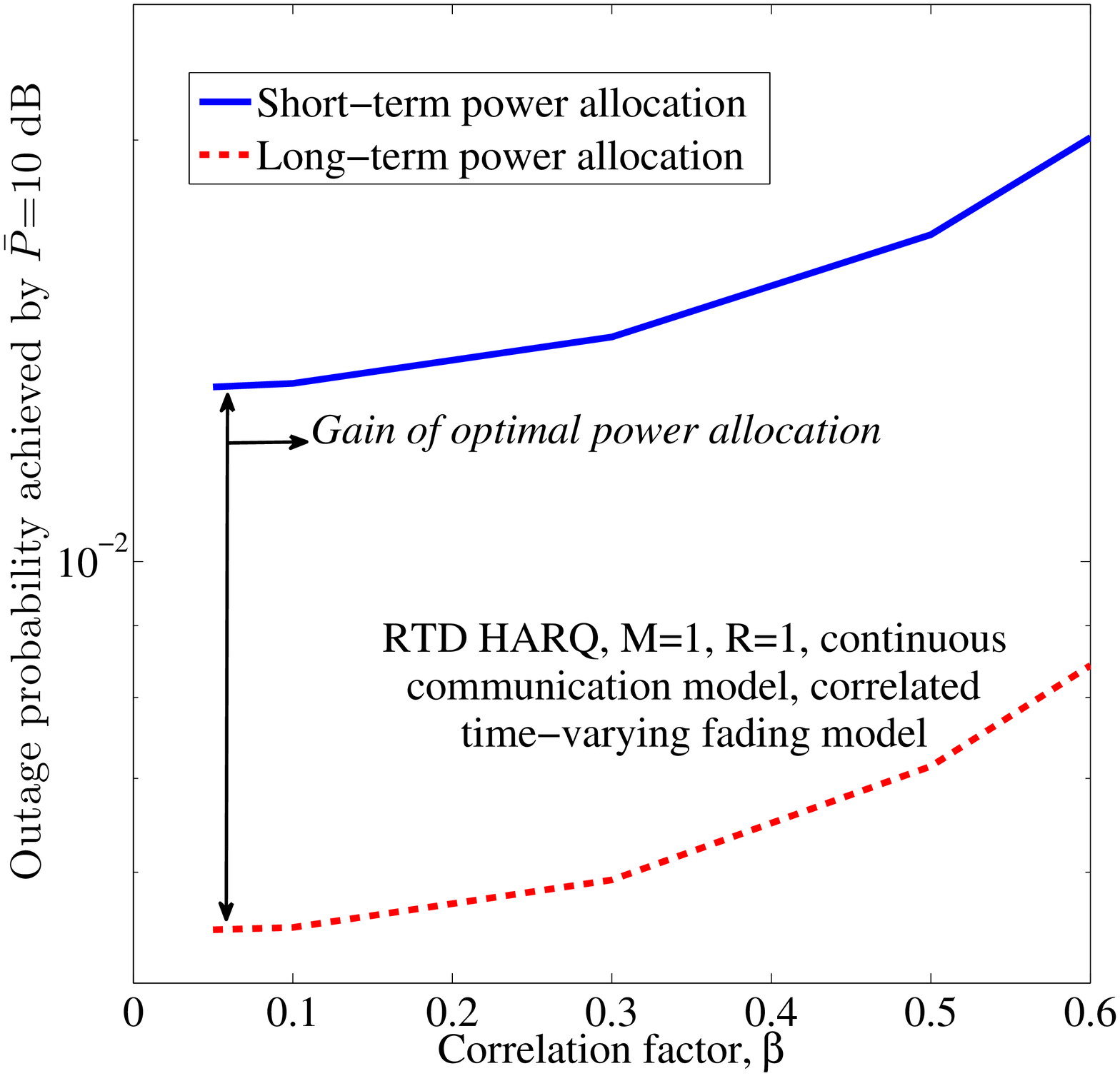}\\\vspace{-4mm}
\caption{Outage probability achieved by average power $\bar P=10$ dB and different correlation factors $\beta$. Correlated Rayleigh fading channel model of \cite{6051400}. The difference between two curves, i.e., the gain of optimal power allocation is almost constant for different values of $\beta$.  }\label{figure111}
\vspace{-4mm}
\end{figure}
\begin{figure}
\vspace{-0mm}
\centering
  \includegraphics[width=0.92\columnwidth]{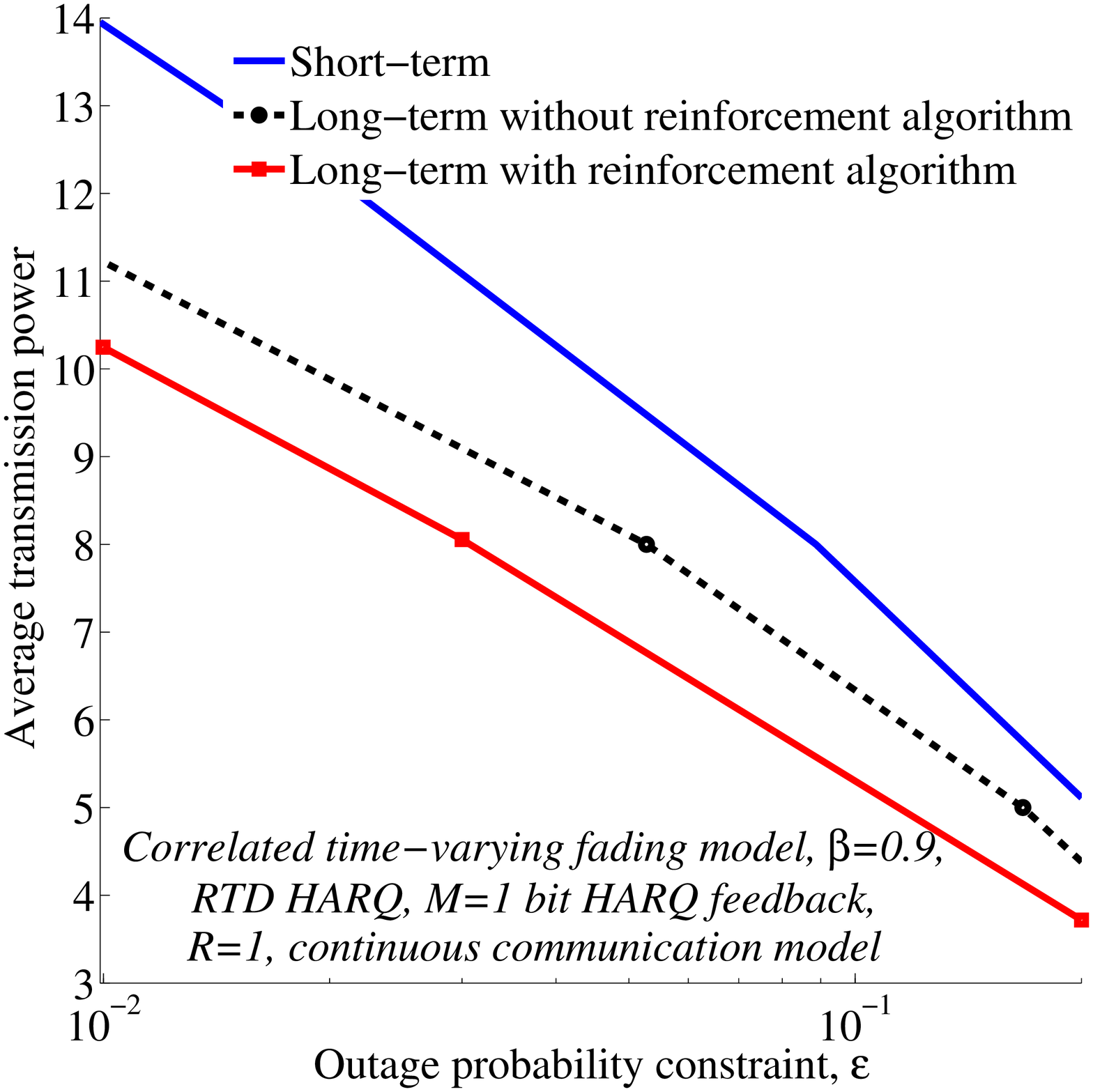}\\\vspace{-4mm}
\caption{Outage-limited average power for different power allocation schemes, RTD HARQ, $M=1$. Correlated Rayleigh fading channel model of \cite{6051400}, $\beta=0.9$. For the reinforcement-based scheme, Algorithm 2 is used where the constants $d_1,\ldots,d_4$ are optimized, in terms of average power, for every given outage probability.  }\label{figure111}
\vspace{-2mm}
\end{figure}

To investigate the effect of reinforcement-based schemes, we obtain the outage-limited average power of an RTD protocol using Algorithm 2. The algorithm follows the same structure as in adaptive quantization schemes \cite[Section 11.6]{khaled}.\footnote{Algorithm 2 is presented for the case with a maximum of $M=1$ retransmission. The extension of the results to the case with arbitrary number of retransmissions is straightforward.} For a given outage probability constraint, the updating steps $d_i,i=1,\ldots,4,$ and the initial transmission power $P_\text{initial}$ of the algorithm are optimized by exhaustive search, to minimize the average power. The simulation results are presented in Fig. 14 which show that, depending on the fading condition, considerable performance improvement can be achieved by applying reinforcement-based techniques. However, as mentioned before, reinforcement-based schemes increase the implementation complexity substantially and are sensitive to feedback channel noise. Also, reinforcement algorithms are not useful in the bursting communication model or in the continuous communication model with uncorrelated fast-fading condition.

\begin{algorithm} [tbh]
\caption{ARQ-based data transmission by a reinforcement algorithm}
\begin{itemize}
\item[I.] For a given initial transmission rate $R$, set the initial transmission power to $\breve{P}=P_\text{initial}.$
\item[II.] Start a packet transmission with power $\breve{P}$ and do the following procedures
\begin{itemize}
  \item[1)] If the codeword is decoded in round $m=1$, $\breve{P}\leftarrow(1-d_1)\breve{P},$ and go to II.
  \item[2)] If the codeword is not decoded in round $m=1$, set $\breve{P}\leftarrow(1+d_2)\breve{P}$ and retransmit the codeword. Continue as follows:
      \begin{itemize}
        \item[a)]  If the codeword is decoded in round $m=2$, $\breve{P}\leftarrow(1-d_3)\breve{P}$ and go to II.
        \item[b)]  If the codeword is not decoded in round $m=2$, declare an outage, set $\breve{P}\leftarrow(1+d_4)\breve{P}$ and go to II.
      \end{itemize}
\end{itemize}

\end{itemize}

\end{algorithm}
\vspace{-0mm}
The intuition behind the better system performance in the continuous communication model (in comparison with the bursting) is that more packets are transmitted in \emph{good} channels in continuous communication than in the bursting model. In other words, the good channels are more efficiently exploited in the continuous model. Finally, it is interesting to note that, comparing with the case when there is no feedback (no retransmission), considerable performance improvement is obtained with limited number of feedback bits (see Figs. 8-10).
\vspace{-5mm}
\section{Conclusion}
Considering delay-sensitive networks, this paper analyzed the performance of the RTD and INR HARQ protocols in outage-limited data transmission conditions. The results emphasize that considerable performance improvement (in terms of both the required transmission power and the system throughput) is achieved with limited number of HARQ-based feedback bits. To minimize the outage-limited average power in the RTD and INR protocols, the transmission powers and energies must increase in every retransmission, respectively. Moreover, for sufficiently large number of retransmissions, the optimal (re)transmission powers can be determined via a specific sequence of numbers. This sequence, which is independent of the fading distribution, converges to a geometric sequence when the order of retransmission rounds increases. Also, adaptive power allocation between the data retransmission rounds is very beneficial in hard outage probability constraints while its effect diminishes in more relaxed outage-limited conditions. Depending on the channel condition, the power efficiency can be improved by reinforcement-based power allocation schemes, at the cost of higher implementation complexity and sensitivity to imperfect feedback signal. Finally, studying the system performance with an imperfect feedback channel and implementation of sub-packeting schemes are interesting extensions of the paper.
\vspace{-3mm}
\appendices
\section{Proof of Theorem 4}
Assuming $R^{(m)}=\frac{R}{m}$, Theorem 3 implies $l_1=l_2$, i.e., the codewords lengths are the same in the INR protocol. Therefore, from (33) and (40), the average transmission power in both schemes can be rewritten as
\vspace{-0mm}
\begin{align}
\varphi  = \frac{{{P_1} + {P_2}{F_G}(\frac{{{e^R} - 1}}{{{P_1}}})}}{{1 + {F_G}(\frac{{{e^R} - 1}}{{{P_1}}})}}.
\end{align}
To prove part (a), we consider two cases $\{\text{Case}\, 1: ({P_1} = P + \Delta ,{P_2} = P)\}$ and $\{\text{Case}\, 2: ({P_1} = P,{P_2} = P + \Delta )\},\, \Delta \ge 0$ and show that less average transmission power is obtained in the second case. Note that based on (10) and (23) there is no preference between the transmission powers from the outage probability constraint perspective, as the powers are interchangeable. Then, based on the following inequalities
\vspace{-0mm}
\begin{align}
\begin{array}{l}
 {\varphi _{\text{case}\,1}} = \frac{{P + \Delta  + P{F_G}(\frac{{{e^R} - 1}}{{P + \Delta }})}}{{1 + {F_G}(\frac{{{e^R} - 1}}{{P + \Delta }})}} \ge \frac{{P + (P + \Delta ){F_G}(\frac{{{e^R} - 1}}{P})}}{{1 + {F_G}(\frac{{{e^R} - 1}}{P})}} = {\varphi _{\text{case}\,2}} \\
  \Leftrightarrow \,(P + \Delta  + P{F_G}(\frac{{{e^R} - 1}}{{P + \Delta }}))(1 + {F_G}(\frac{{{e^R} - 1}}{P})) \\\,\,\,\,\,\,\,\,\,\,\,\,\,\,\,\,\,\,\,\,\,\,\,\,\,\,\,\,\,\,\,\,\ge (P + (P + \Delta ){F_G}(\frac{{{e^R} - 1}}{P}))(1 + {F_G}(\frac{{{e^R} - 1}}{{P + \Delta }})) \\
  \Leftrightarrow 1 \ge {F_G}(\frac{{{e^R} - 1}}{{P + \Delta }}){F_G}(\frac{{{e^R} - 1}}{P}) \\
 \end{array}
\end{align}
it is obvious that less average transmission power is obtained in the second case, i.e., the optimal (re)transmission powers minimizing the average transmission power should be increasing with the number of retransmission rounds.

To prove part (b), we assume that the transmission powers $P_1$ and $P_2$ in (44) have been optimized for the RTD-based scheme, i.e., according to the constraint $\Pr \{ \log (1 + g({P_1} + {P_2})) < R\}  \le \epsilon$. Now, the same transmission power $P_1$ is considered for the first transmission round of the INR-based scheme which is not necessarily optimal for this protocol. Then, according to the property $\log (1 + ax) + \log (1 + by) \ge \log (1 + ax + by),\,\forall x,y,a,b \ge 0$, it can be easily shown that less transmission power than $P_2$ is required to satisfy the INR-based outage probability constraint $\Pr \{ \log (1 + g{P_1}) + \log (1 + g{P_2}) < R\}  \le \epsilon$. That is, defining
\vspace{-0mm}
\begin{align}
w \buildrel\textstyle.\over= \mathop {\arg }\limits_x \{ \Pr \{ \log (1 + g{P_1}) + \log (1 + gx) < R\}  = \epsilon \}\nonumber
\end{align}
we have $w \le P_2$. Therefore, as the average power in both schemes is found by (44), the optimal outage-limited average transmission power required in the INR HARQ scheme is less than the one required in the RTD protocol. Finally, the theorem can be further verified in Figs. 4-7.

\vspace{-0mm}
\bibliographystyle{IEEEtran} 
\bibliography{revisionrevisionmasterICC3}
\vspace{-2mm}
\vfill

\end{document}